\documentclass[prd,nofootinbib,preprint,superscriptaddress]{revtex4-1}
\pdfoutput=1

\usepackage{graphicx}
\usepackage{epsfig}
\usepackage{amsmath}
\usepackage{amsfonts}
\usepackage{amssymb}
\usepackage{color}
\usepackage[bottom]{footmisc}
\usepackage{array,multirow,makecell}
\usepackage{slashed}
\usepackage[colorlinks,citecolor=blue]{hyperref}
\usepackage{float}
\usepackage{ulem}
\usepackage[utf8]{inputenc}
\usepackage[english]{babel}
\usepackage{url}
\usepackage{hyperref}
\usepackage{xcolor}
\usepackage{subfigure}
\usepackage{lipsum}
\usepackage[bottom]{footmisc}
\usepackage{pifont}

\usepackage[most]{tcolorbox}
\newtcbox{\mymath}[1][]{%
    nobeforeafter, math upper, tcbox raise base,
    enhanced, colframe=blue!30!black,
    colback=blue!30, boxrule=1pt,
    #1}

\usepackage{blindtext}

\def\lsim{\raise0.3ex\hbox{$\;<$\kern-0.75em\raise-1.1ex\hbox{$\sim\;$}}}
\def\gsim{\raise0.3ex\hbox{$\;>$\kern-0.75em\raise-1.1ex\hbox{$\sim\;$}}}

\def\bea{\begin{eqnarray}}
\def\eea{\end{eqnarray}}

\newcommand{\mdm}{m_\psi}
\newcommand{\Tbh}{T_\text{BH}}
\newcommand{\Tin}{T_\text{in}}
\newcommand{\Tev}{T_\text{evap}}

\newcommand{\Min}{m_\text{in}}
\newcommand{\Trh}{T_\text{rh}}
\newcommand{\Tmax}{T_\text{max}}
\newcommand{\ogw}{\Omega_\text{gw}}
\newcommand{\gss}{g_{\star,s}}
\newcommand{\gsH}{g_{\star,H}}

\begin{document}
\title{Gravitational wave signatures of PBH-generated \\ baryon-dark matter coincidence}
\author{Basabendu Barman}
\email{basabendu88barman@gmail.com}
\affiliation{Institute of Theoretical Physics, Faculty of Physics, University of Warsaw,\\ ul. Pasteura 5, 02-093 Warsaw, Poland}

\author{Debasish Borah}
\email{dborah@iitg.ac.in}
\affiliation{Department of Physics, Indian Institute of Technology Guwahati, Assam 781039, India}

\author{Suruj Jyoti Das}
\email{suruj@iitg.ac.in}
\affiliation{Department of Physics, Indian Institute of Technology Guwahati, Assam 781039, India}

\author{Rishav Roshan}
\email{rishav.roshan@gmail.com}
\affiliation{Department of Physics, Kyungpook National University, Daegu 41566, Korea}
\begin{abstract}
We propose a new way of probing non-thermal origin of baryon asymmetry of universe (BAU) and dark matter (DM) from evaporating primordial black holes (PBH) via stochastic gravitational waves (GW) emitted due to PBH density fluctuations. We adopt a baryogenesis setup where CP violating out-of-equilibrium decays of a coloured scalar, produced non-thermally at late epochs from PBH evaporation, lead to the generation of BAU. The same PBH evaporation is also responsible for non-thermal origin of superheavy DM. Unlike the case of  baryogenesis {\it via leptogeneis} that necessarily corners the PBH mass to $\sim\mathcal{O}(1)$ g, here we can have PBH mass as large as $\sim\mathcal{O}(10^7)$ g due to the possibility of producing BAU directly below sphaleron decoupling temperature. Due to the larger allowed PBH mass we can also have observable GW with mHz-kHz frequencies originating from PBH density fluctuations keeping the model constrained and verifiable at ongoing as well as near future GW experiments like LIGO, BBO, DECIGO, CE, ET etc. Due to the presence of new coloured particles and baryon number violation, the model also has complementary detection prospects at laboratory experiments.
\end{abstract}
\maketitle
{
  \hypersetup{linkcolor=black}
  \tableofcontents
}
\section{Introduction}
\label{sec:intro}
Observational evidence from astrophysics and cosmology based experiments suggest that approximately $27\%$ of present universe's energy density is composed of dark matter (DM). More quantitatively, DM abundance is reported in terms of density parameter $\Omega_{\rm DM}$ and reduced Hubble constant $h = \text{Hubble Parameter}/(100 \;\text{km} ~\text{s}^{-1}
\text{Mpc}^{-1})$ as \cite{Aghanim:2018eyx}
\begin{equation}
\Omega_{\text{DM}} h^2 = 0.120\pm 0.001
\label{dm_relic}
\end{equation}
\noindent at 68\% CL. While around $5\%$ of the present universe is composed of ordinary matter or baryons, the highly asymmetry nature of it leads to the longstanding puzzle of baryon asymmetry of universe (BAU). This observed excess of baryons over anti-baryons is quantified in terms of the baryon to photon ratio as \cite{Aghanim:2018eyx} 
\begin{equation}
\eta_B = \frac{n_{B}-n_{\overline{B}}}{n_{\gamma}} \simeq 6.2 \times 10^{-10}, 
\label{etaBobs}
\end{equation} 
based on the cosmic microwave background (CMB) measurements which also agrees well with the big bang nucleosynthesis (BBN) estimates \cite{Zyla:2020zbs}. While the origin of DM as well as BAU remain unexplained within the framework of the standard model (SM), the resemblance between their abundances namely, $\Omega_{\rm DM} \approx 5\,\Omega_{\rm baryon}$ might deserve an explanation based on dynamical origin. Such cogenesis mechanisms for DM and BAU\footnote{A brief review of such cogenesis can be found in \cite{Boucenna:2013wba}.} are broadly classified into two categories namely the asymmetric dark matter (ADM) scenario \cite{Nussinov:1985xr, Davoudiasl:2012uw, Petraki:2013wwa, Zurek:2013wia, DuttaBanik:2020vfr, Barman:2021ost, Cui:2020dly} and the WIMPy baryogenesis scenario \cite{Cui:2011ab, Bernal:2012gv, Bernal:2013bga, Kumar:2013uca, Racker:2014uga, Dasgupta:2016odo, Borah:2018uci, Borah:2019epq, Dasgupta:2019lha, Mahanta:2022gsi} where the asymmetry in either baryon or lepton sector is produced from particle DM annihilations \cite{Yoshimura:1978ex, Barr:1979wb, Baldes:2014gca}.

In spite of the theoretical appeal in many of these beyond standard model (BSM) frameworks for cogenesis, experimental verification has been a challenge due to persistent lack of any signatures either at DM search experiments \cite{LZ:2022ufs} or at other experiments like the large hadron collider (LHC). This has motivated the search for complementary signatures like stochastic gravitational wave (GW) background. For example, GW signatures of leptogenesis has been discussed in \cite{Dror:2019syi, Blasi:2020wpy, Fornal:2020esl, Samanta:2020cdk, Datta:2022tab, Datta:2020bht, Samanta:2021zzk, Barman:2022yos, Ghoshal:2022kqp, Dasgupta:2022isg, Borah:2022cdx, Lazarides:2022ezc}. Similarly, some recent studies on GW signatures of different DM scenarios can be found in \cite{Yuan:2021ebu, Tsukada:2020lgt, Chatrchyan:2020pzh, Bian:2021vmi, Samanta:2021mdm, Borah:2022byb, Azatov:2021ifm, Azatov:2022tii, Baldes:2022oev,Chakrabortty:2020otp, Borah:2022vsu, Lazarides:2022spe, Costa:2022oaa, Ghosh:2022fzp}. In this work, we consider a cogenesis setup where ultralight primordial black holes (PBH) can play a crucial role. The role of PBH evaporation on baryogenesis was first pointed out in \cite{Hawking:1974rv, Carr:1976zz} and has been studied subsequently by several authors in different contexts~\cite{Baumann:2007yr, Hook:2014mla, Hamada:2016jnq, Morrison:2018xla, Hooper:2020otu, Perez-Gonzalez:2020vnz, Smyth:2021lkn, Bernal:2022pue, Ambrosone:2021lsx}. Common origin of baryogenesis via leptogenesis \cite{Fukugita:1986hr} and DM from evaporating PBH has been studied in several works \cite{Fujita:2014hha, Datta:2020bht, JyotiDas:2021shi,Barman:2021ost, Barman:2022gjo}. Such PBH-generated cogenesis mechanisms have very interesting GW signatures. For example, evaporating ultra-light PBH can lead to high frequency stochastic GW generation \cite{Anantua:2008am, Dolgov:2011cq}. On the other hand, the PBH formation mechanism can itself lead to induced GW \cite{Saito:2008jc, Domenech:2021ztg, Kohri:2018awv}. Recently, the authors of \cite{Bhaumik:2022pil} also proposed a doubly peaked GW spectrum in PBH-generated baryogenesis scenarios by considering first order adiabatic perturbation from inflation and the isocurvature perturbations due to PBH distribution to source the tensor perturbations in second-order. In \cite{Gehrman:2022imk}, authors considered high frequency GW signatures in a baryon-DM cogenesis setup from PBH by considering the scalar perturbations responsible for PBH formation. In another recent work \cite{Borah:2022imk}, a high scale leptogenesis and superheavy DM scenario in the presence of PBH was shown to produce unique GW spectrum via PBH density fluctuations \cite{Papanikolaou:2020qtd, Domenech:2021wkk, Domenech:2020ssp, Domenech:2021ztg}.

In this work, we consider a minimal particle physics setup to realize baryogenesis due to out-of-equilibrium CP violating decay of a coloured scalar. The baryon asymmetry can be generated at low scale, even below the sphaleron decoupling temperature due to non-thermal origin of the coloured scalar from PBH evaporation. Due to the possibility of low scale generation of baryon asymmetry, the allowed PBH mass range can be much bigger compared to the one in several PBH assisted leptogenesis works discussed in the literature. This also brings the GW spectrum created by PBH density fluctuations to the observable ballpark in mHz-kHz frequencies with peak amplitudes lying within reach of even LIGO as well as several planned experiments. While such PBH dominated epoch in the early universe typically leads to the over-production of DM except for superheavy regime, we find interesting correlations between DM mass and the coloured scalar mass responsible for creating BAU. Our analysis also differs from the recent work~\cite{Gehrman:2022imk} where a second population of heavy stable PBH was considered which itself act as  DM, in addition to the difference in GW production source and frequencies (MHz-GHz ballpark). While our GW signatures are similar to the recent work~\cite{Borah:2022imk}, there exist sharp differences in the cogenesis mechanism. In~\cite{Borah:2022imk}, high scale thermal leptogenesis leads to over-production of baryon asymmetry which gets diluted due to PBH evaporation at late epochs which also leads to the production of superheavy DM. On the other hand, we are studying direct baryogenesis (without taking the leptogenesis route) and DM production from PBH evaporation. The complementary detection prospects are more promising in the present setup due to the presence of additional coloured particles and baryon number violation. Depending upon the mass of the coloured scalar and its couplings, the model can also have complementary detection prospects at the LHC and experiments looking for baryon number violating processes like neutron-antineutron $(n-\bar{n})$ oscillations. Our aim here is to provide a testable scenario, based on existing particle physics model, that is capable of simultaneously explaining baryonic and dark matter abundance, thanks to the underlying gravitational production associated with PBH evapoaration.

This paper is organized as follows. In section \ref{sec:framework} we briefly discuss the model and source of baryon asymmetry. In section \ref{sec:pbh}, we review the role of PBH in baryogenesis while also commenting upon difference from PBH assisted leptogenesis followed by discussion of the results related to BAU and DM from PBH. In section \ref{sec:GW}, we discuss the details of GW generation in our setup and finally conclude in section \ref{sec:concl}.

\section{The Framework}
\label{sec:framework}
\subsection{Field content and interactions}
In this section we describe the underlying particle physics model responsible for baryogenesis. To realise baryogenesis, we follow a simple particle physics setup similar to the earlier works~\cite{Allahverdi:2010im, Allahverdi:2010rh,Allahverdi:2013tca, Allahverdi:2013mza,Allahverdi:2017edd}. We consider renormalisable baryon number (B) violating terms involving newly introduced particles as a source of BAU. In order to avoid $\Delta B=1$ processes leading to proton decay, we consider $\Delta B=2$ processes having other observable signatures like $n-\bar{n}$ oscillations \cite{Baldo-Ceolin:1994hzw, Super-Kamiokande:2011idx, SNO:2017pha}. We consider the presence of two iso-singlet scalars $S_i$, with $i=1,\,2$, that transform as $SU(3)_c$ triplets under the SM. The scalars also carry non-trivial $U(1)_Y$ charges that allow us to have a direct coupling term to right handed down type quarks as $S\,d^c\,d^c$. The presence of at least two $S$ is needed to produce a baryon asymmetry from the interference of tree and one-loop diagrams in a decay process governed by the $S\,d^c\,d^c$ interactions. However, although necessary, this is not sufficient. The reason being that the total asymmetry vanishes after summing over all flavors of $d^c$ in the final and intermediate states \cite{Kolb:1979qa}. One therefore, requires additional baryon number violating interactions, and the simplest renormalisable term as such is $S \psi\,u^c$ where $\psi$ is a SM gauge singlet field. The newly added fields and their corresponding charges under the SM gauge symmetry are listed in Table~\ref{tab:fields}. With this particle content at our disposal, we can now write the relevant part of the renormalisable Lagrangian as \cite{Allahverdi:2013mza, Allahverdi:2017edd, Dev:2015uca, Davoudiasl:2015jja}
\begin{align}\label{eq:lg}
& -\mathcal{L}\supset \lambda\,S\,\psi\,u^c + \lambda'\,S^\star\,d^c\,d^c + \frac{1}{2}\,\mdm\,\overline{\psi^c}\,\psi+\text{h.c.}\,,
\end{align}
where we have suppressed all the flavour indices. Clearly, the Majorana mass term of $\psi$ is the source of baryon number violation ($\Delta B=2$) in this model.
\begin{table}[htb!]
\centering
\begin{tabular}{|c|c|c|c|c|c|c|}
\hline
Fields &$SU(3)_c$ & $SU(2)_L$ & $U(1)_Y$ \\
\hline\hline
$u^c$ & 3 & 1 & $-4/3$ \\
$d^c$ & 3 & 1 & $+2/3$ \\
\hline
$S_i\,(i\in 1,2)$ & 3 & 1 &  $+4/3$ \\
$\psi$ & 1 & 1 &  0 \\
\hline\hline
\end{tabular}
\caption{Relevant fields including the newly added ones and their corresponding charges under the SM gauge symmetry.}
\label{tab:fields}
\end{table}

The above particle content does not lead to a naturally stable DM candidate. For the interaction Lagrangian given in Eq. \eqref{eq:lg}, the neutral singlet fermion $\psi$ can be made kinematically stable by choosing its mass as $m_p-m_e \leq m_\psi \leq m_p+m_e$, forbidding both the possibilities of proton and $\psi$ decays. However, in the absence of additional symmetries, $\psi$ can couple to lepton and Higgs doublets of the SM opening up another decay channel. Even if additional symmetry is introduced to forbid such decay into leptons, a PBH dominated phase typically over-produces DM with mass $\mdm\sim 1$ GeV, as we discuss below. Therefore, in our minimal setup, $\psi$ can not be a DM candidate. Therefore, $\psi$ can be much heavier in our setup while obeying the lower bound $m_p-m_e \leq m_\psi$ to forbid the corresponding proton decay mode $p \rightarrow \psi \, e^+ \, \nu_e$. Interestingly, multiple generations of $\psi$ can also play a role in generating light neutrino mass via type-I seesaw mechanism \cite{Minkowski:1977sc,Yanagida:1979as,Gell-Mann:1979vob,Mohapatra:1980yp}. We consider DM to be purely gravitational such that it is produced dominantly from PBH evaporation. Since both DM and the coloured scalar S responsible for BAU originate non-thermally from PBH evaporation, this leads to a strong correlation among DM, coloured scalar as well as PBH initial masses. In addition, a purely gravitational DM is also motivated from the fact that all observational evidences of DM are based on its gravitational interactions only, with direct detection experiments continuing their null results.

\subsection{Generation of baryon asymmetry}
Baryon asymmetry is generated through the decay of the coloured scalars $S_i$, leading to CP asymmetries, via interference between a tree and loop-level diagrams as in Ref.~\cite{Allahverdi:2017edd} 
\begin{align}\label{eq:cp-asym}
&  \epsilon_\alpha =
\frac{1}{8\,\pi}\,\frac{\sum_{i,j,k}\,\text{Im}\,\left(\lambda^*_{\alpha k}\lambda_{\beta k}\lambda^{\prime *}_{\alpha  ij}\lambda^{\prime}_{\beta ij}\right)}{\sum_{i,j}|\lambda^{\prime}_{\alpha ij}|^2 + \sum_i |\lambda_{\alpha i}|^2}
\times\frac{\left(m_{S_\alpha}^2-m_{S_\beta}^2\right)\,m_{S_\alpha}\,m_{S_\beta}}{\left(m_{S_\alpha}^2-m_{S_\beta}^2\right)^2+m_{S_\alpha}^2\,\Gamma_{S_\beta}^2}\,;\text{with}\,\alpha\,,\beta=1\,,2\,;\,\alpha\neq\beta\,,
\end{align}
such that the final baryon asymmetry, in the non-thermal ballpark (where washouts are negligible) can be estimated as
\begin{align}\label{eq:etaB}
& Y_B = \frac{n_B}{s} = \epsilon_1\,\frac{n_{S_1}}{s} + \epsilon_2\,\frac{n_{S_2}}{s}.
\end{align}
In Eq. \eqref{eq:cp-asym}, the decay width of $S_\alpha$, at tree level, is estimated as
\begin{align}\label{eq:S-decay}
& \Gamma_{S_\alpha} = \frac{m_{S_\alpha}}{16\pi}\,\left[\sum_i\left|\lambda_{\alpha i}\right|^2+\sum_{i,j}\left|\lambda_{\alpha ij}'\right|^2\right]\,.
\end{align}
In order to compute the CP asymmetries from Eq.~\eqref{eq:cp-asym}, we parameterize the Yukawa matrices in Eq.~\eqref{eq:lg} as
\begin{equation}
\lambda=\left(
\begin{array}{ccc}
 \lambda & \lambda &  \lambda \\
 \lambda e^{i\phi_1} & \lambda e^{i\phi_1} & \lambda e^{i\phi_1} \\
\end{array}
\right)\,,\lambda_1^{\prime}=\left(
\begin{array}{ccc}
0 & \lambda^{\prime} &  \lambda^{\prime} \\
 \lambda^{\prime} & 0 & \lambda^{\prime} \\
 \lambda^{\prime} & \lambda^{\prime} & 0 \\
\end{array}
\right)\,,\lambda_2^{\prime}=\left(
\begin{array}{ccc}
 0 & \lambda^{\prime} e^{i\phi_2} &  \lambda^{\prime} e^{i\phi_2} \\
\lambda^{\prime} e^{i\phi_2} & 0 & \lambda^{\prime} e^{i\phi_2} \\
\lambda^{\prime} e^{i\phi_2} & \lambda^{\prime} e^{i\phi_2} & 0 \\
\end{array}
\right)\,,
\end{equation}
which provides us with two more free parameters, namely the arbitrary CP phases $\phi_{1,2}$. Also, the diagonal entries of $\lambda^\prime_{1,2}$ are zero due to the requirement of colour anti-symmetry. Note that $\lambda^\prime_{\alpha ij}$ denotes the components of the matrices $\lambda^\prime_\alpha$ with $\alpha=1,2$. Utilizing this parametrisation, from Eq.~\eqref{eq:cp-asym} we find $\epsilon_1+\epsilon_2 = 0.1$ for $\lambda\,,\lambda'\simeq 10^{-3}$ with $m_{S_1}\approx m_{S_2}$. Typically, this corresponds to {\it resonant} baryogenesis scenario, where  the mass difference between $S_{1,2}$ is of the same order as their decay width: $\Delta m\equiv |m_{S_1}- m_{S_2}|\sim \Gamma_{S_{1,2}}/2\ll m_{S_{1,2}}$ \cite{Allahverdi:2017edd}, similar to the resonant leptogenesis framework \cite{Pilaftsis:2003gt}. Since $\Gamma_S\propto m_S$, hence the CP asymmetry in the resonance limit becomes independent of the scalar mass.

\subsection{Constraints from the LHC and $n-\bar{n}$ Oscillations}
While we do not perform a detailed collider study of our model here, we briefly mention the existing bounds on coloured scalars, which most part of our parameter space satisfy. The on-shell production of $S$ via $pp\to S\to d^c_id^c_j$, can lead to dijet resonance that in turn can constraint the mass of $S$ and the couplings $\lambda\,,\lambda'$, as discussed in \cite{Allahverdi:2017edd}. A search for narrow resonances decaying to dijet final states put upper limits at 95\% confidence level on the production cross section with masses above 0.6 TeV at $\sqrt{s}=13$ TeV, with an integrated luminosity of $12.9\,\text{fb}^{-1}$~\cite{CMS:2016gsl}. This excludes colour octet scalars below 3 TeV. This bound has been updated for an integrated luminosity of $137\,\text{fb}^{-1}$, that searches for a narrow (or broad) s-channel dijet resonance with mass above 1.8 TeV, constraining colour octet scalars below 3.7 TeV~\cite{CMS-PAS-EXO-19-012}. There is also a possible monojet signal of this model at the LHC from the on-shell production of $S$ and its subsequent decay through: $pp\to S\to \psi u^c$, that can be constrained from monojet plus missing energy searches from CMS~\cite{CMS:2021far} and ATLAS~\cite{ATLAS:2021kxv}. On the other hand, the colored scalars can also be pair-produced at a hadron collider: $pp\to SS^\star$, purely through QCD interactions, independent of the Yukawa couplings. The subsequent decay of $S\to d^c_i\,d^c_j$ will then lead to paired dijet resonance. As explained in~\cite{Allahverdi:2017edd}, the baryon number violating term in the Lagrangian can induce a $B$-violating dimension-9 operator of the form $\left(m_\psi/m_S^6\right)\,(\lambda^\frac{1}{2}\,\lambda')^4\,\left(u^c\,u^c\right)\,\left(d^c\,b^c\right)\,\left(d^c\,b^c\right)$, corresponding to $\Delta B=2$ transition, that can lead to $n-\bar{n}$ oscillation with the amplitude
\begin{align}\label{eq:nnbar}
\mathcal{M}_{n\bar{n}}\simeq  \frac{\lambda^2\,(\lambda')^4\,m_\psi}{8\,\pi^2\,m_S^6}\,\ln\left[\frac{m_S}{m_\psi}\right]\,,   
\end{align}
at one loop level, which is constrained from current experimental lower bound~\cite{Baldo-Ceolin:1994hzw,Mohapatra:2009wp, Super-Kamiokande:2011idx, SNO:2017pha}: $\mathcal{M}_{n\bar{n}}\leq 10^{-28}\,\text{GeV}^{-5}$ or equivalently, an oscillation lifetime of $\tau_{n\bar{n}}\gtrsim 10^8$ s.

\section{Analytical treatment of PBH assisted cogenesis}
\label{sec:pbh}
Primordial black holes (PBH), originally proposed by Hawking \cite{Hawking:1974rv, Hawking:1975vcx}, can have very interesting cosmological signatures \cite{Chapline:1975ojl, Carr:1976zz} (a recent review of PBH may be found in \cite{Carr:2020gox}). While PBH with suitable mass range can itself be a DM candidate, we are interested in its ultra-light mass regime where it is not long-lived enough to be DM but can play non-trivial role in production of DM as well as baryon asymmetry. Role of PBH evaporation on DM genesis has been studied in several works including~\cite{ Gondolo:2020uqv, Bernal:2020bjf, Green:1999yh, Khlopov:2004tn, Dai:2009hx, Allahverdi:2017sks, Lennon:2017tqq, Hooper:2019gtx, Sandick:2021gew}. Similarly, the role of PBH evaporation on baryogenesis was first pointed out in \cite{Hawking:1974rv, Carr:1976zz} and has been studied subsequently by several authors in different contexts~\cite{Baumann:2007yr, Hook:2014mla, Fujita:2014hha, Hamada:2016jnq, Morrison:2018xla, Hooper:2020otu, Perez-Gonzalez:2020vnz, Datta:2020bht, JyotiDas:2021shi, Smyth:2021lkn, Barman:2021ost, Bernal:2022pue, Ambrosone:2021lsx}. 

We assume PBHs to have formed after inflation during the era of radiation domination. Assuming radiation domination, the mass of the black hole from gravitational collapse is typically close to the value enclosed by the post-inflation particle horizon and is given by~\cite{Fujita:2014hha,Masina:2020xhk}
\begin{equation}
m_\text{BH}^{\rm in}=\frac{4}{3}\,\pi\,\gamma\,\Bigl(\frac{1}{H\left(T_\text{in}\right)}\Bigr)^3\,\rho_\text{rad}\left(T_\text{in}\right)\,,
\label{eq:pbh-mass}
\end{equation}
with 
\begin{equation}
\rho_\text{rad}\left(T_\text{in}\right)=\frac{3}{8\,\pi}\,H\left(T_\text{in}\right)^2\,M_P^2\,,
\end{equation}
where $H$ is the Hubble parameter, $M_P$ is the Planck mass and $\gamma\simeq 0.2$ is a numerical factor which contains the uncertainty of the PBH formation.  As mentioned earlier,  PBHs are produced during the radiation dominated epoch, when the SM plasma has a temperature $T=T_\text{in}$ which is given by
\begin{equation}
T_\text{in}=\Biggl(\frac{45\,\gamma^2}{16\,\pi^3\,g_\star\left(T_\text{in}\right)}\Biggr)^{1/4}\,\sqrt{\frac{M_P}{m_\text{BH}(T_\text{in})}}\,M_P\,.
\label{eq:pbh-in}
\end{equation}
Once formed, PBH can evaporate by emitting Hawking radiation \cite{Hawking:1974rv, Hawking:1975vcx}. A PBH can evaporate efficiently into particles lighter than its instantaneous Hawking temperature $T_\text{BH}$ defined as \cite{Hawking:1975vcx}
\begin{equation}
T_{\rm BH}=\frac{1}{8\pi\,G\, m_{\rm BH}}\approx 1.06~\left(\frac{10^{13}\; {\rm g}}{m_{\rm BH}}\right)~{\rm GeV}\,,
\end{equation}
\noindent where $G$ is the universal gravitational constant. The mass loss rate can be parametrised as \cite{MacGibbon:1991tj}
\begin{equation}
\frac{dm_\text{BH}(t)}{dt}=-\frac{\mathcal{G}\,g_\star\left(T_\text{BH}\right)}{30720\,\pi}\,\frac{M_P^4}{ {m_\text{BH}(t)^2}}\,,
\label{eq:pbh-dmdt}
\end{equation}
where $\mathcal{G}\sim 4$ is the grey-body factor. Here we ignore the temperature dependence of $g_\star$ during PBH evolution, valid in the era prior to sphaleron decoupling temperature. On integrating Eq.~\eqref{eq:pbh-dmdt} we end up with the PBH mass evolution equation as
\begin{equation}
m_\text{BH}(t)=m_\text{BH}(T_\text{in})\Bigl(1-\frac{t-t_\text{in}}{\tau}\Bigr)^{1/3}\,,
\end{equation}
with 
\begin{equation}
\tau = \frac{10240\,\pi\,(m_\text{BH}^{\rm in})^3}{\mathcal{G}\,g_\star(T_\text{BH})\,M_P^4}\,,
\end{equation}
as the PBH lifetime, and $t_\text{in}$ corresponds to the time at formation of the PBH. Here onward we will use $m^\text{in}_{\rm BH}(T_\text{in})$ simply as $m_\text{in}$. The evaporation temperature can then be computed taking into account $H(T_\text{evap})\sim\frac{1}{\tau^2}\sim\rho_\text{rad}(T_\text{evap})$ as
\begin{equation}
T_\text{evap}\equiv\Bigl(\frac{45\,M_P^2}{16\,\pi^3\,g_\star\left(T_\text{evap}\right)\,\tau^2}\Bigr)^{1/4}\,.
\label{eq:pbh-Tev}
\end{equation}
However, if the PBH component dominates the total energy density of the universe at some epoch, the SM temperature just after the complete evaporation of PBHs is: $\overline{T}_\text{evap}=2/\sqrt{3}\,T_\text{evap}$~\cite{Bernal:2020bjf}. 

The initial PBH abundance is characterized by the dimensionless parameter $\beta$ that is defined as
\begin{equation}
\beta\equiv\frac{\rho_\text{BH}\left(T_\text{in}\right)}{\rho_\text{rad}\left(T_\text{in}\right)}\,,
\end{equation}
corresponding to the ratio of the PBH energy density to the SM energy density at the epoch of PBH formation. Note that, $\beta$ steadily grows until PBH evaporation since the PBH energy density scales like non-relativistic matter $\sim a^{-3}$, while the radiation energy density scales as $\sim a^{-4}$. Therefore, an initially radiation-dominated universe will eventually become matter-dominated if the PBHs are still around. The condition of PBH evanescence during radiation domination can be expressed as~\cite{Masina:2020xhk}
\begin{equation}
\beta<\beta_\text{crit}\equiv \gamma^{-1/2}\,\sqrt{\frac{\mathcal{G}\,g_\star(T_\text{BH})}{10640\,\pi}}\,\frac{M_P}{m_\text{in}}\,,
\label{eq:pbh-ev-rad}
\end{equation}
where $\beta_c \equiv \beta_{\rm crit}$ is the critical PBH abundance that leads to early matter-dominated era. Depending on the initial PBH mass $m_{\text{in}}$, $\beta$ is bounded from below from the requirement of PBH domination, as well as from the above such that the induced GWs do not exceed the limits on the abundance of radiation during BBN. These bounds can be translated into~\cite{Domenech:2020ssp}  \begin{equation}\label{eq:beta}
    1.1 \times 10^{-6} \left(\frac{m_{\rm in}}{10^{4}g}\right)^{-17/24} \gtrsim \beta \gtrsim 6.4 \times 10^{-10} \left(\frac{m_{\rm in}}{10^{4}g}\right)^{-1}\,.
\end{equation}

Since PBH evaporation produces all particles, including radiation that can disturb the successful predictions of BBN, hence we require $T_\text{evap}>T_\text{BBN}\simeq 4$ MeV. This can be translated into an upper bound on the PBH mass. On the other hand, a lower bound on PBH mass can be obtained from the CMB bound on the scale of inflation \cite{Planck:2018jri} : $H_I\equiv H(T_\text{in})\leq 2.5\times 10^{-5}\,M_P$, where $H(T_\text{in})=\frac{1}{2\,t_\text{in}}$ with $t(T_\text{in})=\frac{m_\text{in}}{M_P^2\,\gamma}$ (as obtained from Eq.~\eqref{eq:pbh-mass}). Using these BBN and CMB bounds together, we have a window\footnote{The range of PBH masses between these bounds is at present generically unconstrained~\cite{Carr:2020gox}.} for allowed initial mass for ultra-light PBH that reads $0.1\,\text{g}\lesssim m_\text{in}\lesssim 3.4\times 10^8\,\text{g}$. For simplicity, we consider a monochromatic mass function of PBHs implying all PBHs to have identical masses. Additionally, the PBHs are assumed to be of Schwarzschild type without any spin and charge. Now, the number of any particle $X$ with mass $m_X$ radiated during the evaporation of a single PBH
\begin{equation}
    \mathcal{N}_X = \frac{g_{X,H}}{g_{\star,H}(T_\text{BH})}
    \begin{cases}
       \frac{4\,\pi}{3}\,\Bigl(\frac{m_\text{in}}{M_P}\Bigr)^2 &\text{for } m_X < T_\text{BH}^\text{in}\,,\\[8pt]
        \frac{1}{48\,\pi}\,\Bigl(\frac{M_P}{m_X}\Bigr)^2 &\text{for } m_X > T_\text{BH}^\text{in}\,,
    \end{cases}\label{eq:pbh-num}\,,
\end{equation}
where
\begin{equation}
g_{\star,H}(T_\text{BH})\equiv\sum_i\omega_i\,g_{i,H}\,; g_{i,H}=
    \begin{cases}
        1.82
        &\text{for }s=0\,,\\
        1.0
        &\text{for }s=1/2\,,\\
        0.41
        &\text{for }s=1\,,\\
        0.05
        &\text{for }s=2\,,\\
    \end{cases}
\end{equation}
with $\omega_i=2\,s_i+1$ for massive particles of spin $s_i$, $\omega_i=2$ for massless species with $s_i>0$ and $\omega_i=1$ for $s_i=0$. At temperatures $T_\text{BH}\gg T_\text{EW}\simeq 160$ GeV, PBH evaporation emits the full SM particle spectrum according to their $g_{\star,H}$ weights, while at temperatures below the MeV scale, only photons and neutrinos are emitted. For $T_\text{BH}\gg 100$ GeV (corresponding to $m_\text{BH}\ll 10^{11}$ g), the particle content of the SM corresponds to $g_{\star,H}\simeq 108$. 

\subsection{Leptogenesis from PBH}
\label{sec:lepto}
While our work is related to non-thermal baryogenesis, we briefly comment upon PBH assisted non-thermal leptogenesis, in order to show the key differences in terms of PBH mass. Baryogenesis via leptogenesis \cite{Fukugita:1986hr} is one of the most popular frameworks which can explain the origin of BAU. In such a scenario, a non-zero asymmetry is first created in the lepton sector, and then gets converted into baryon asymmetry via $(B+L)$-violating electroweak sphaleron transitions~\cite{Kuzmin:1985mm}. The simplest leptogenesis scenario involves the inclusion of three heavy right handed neutrinos (RHN) which also takes part in Type-I seesaw mechanism \cite{Mohapatra:1979ia,Yanagida:1979as,GellMann:1980vs,Glashow:1979nm} of neutrino mass generation. 

If we consider non-thermal leptogenesis in the presence of PBH, these RHNs will be dominantly produced from PBH evaporation. The RHNs emitted during PBH evaporation can undergo CP-violating decays, generating lepton asymmetry. It is possible to analytically derive the mass range of RHNs (and PBH) emitted from PBH evaporation that can provide the observed baryon asymmetry. In the Type-I seesaw mechanism, the CP asymmetry parameter $\epsilon$ has an upper bound~\cite{Davidson:2002qv}
\begin{equation}\label{eq:cp-asymDI}
\epsilon\lesssim \frac{3}{16\,\pi}\,\frac{M_1\,m_{\nu,\text{max}}}{v^2}\,,
\end{equation}
where $v=246$ GeV is the SM Higgs VEV and $m_{\nu,\text{max}}$ is the mass of the heaviest light neutrino. On the other hand, the final asymmetry generated by RHNs produced from PBH evaporation is given by $Y_B^\text{obs}=n_B/s\Big|_{T_0}=\frac{1}{\zeta}\,\mathcal{N}_X\,\epsilon\,a_\text{sph}\,Y_\text{PBH}\Big|_{\Tev}\simeq 8.7\times 10^{-11}$~\cite{Aghanim:2018eyx}, where $a_\text{sph}\simeq 1/3$ and $T_0$ is the present temperature of the universe. These together constrain the mass of the RHN  produced from PBH evaporation {{during PBH domination},} both from above and from below~\cite{Fujita:2014hha,Datta:2020bht,Barman:2021ost,Barman:2022gjo}
\begin{equation}
    M_1 
    \begin{cases}
        > \frac{4\,g_{\star,H}(\Tin)}{g_{X,H}\,a_\text{sph}}\,\zeta\,\frac{Y_B^0}{Y_\text{PBH}^\text{evap}}\,\frac{v^2\,M_P^2}{m_\nu\,\Min^2} &\text{for } M_1 < \Tbh^\text{in}\,;\\[8pt]
        < \frac{g_{X,H}\,a_\text{sph}}{256\,\pi^2\,g_{\star,H}}\,\frac{1}{\zeta}\,\frac{Y_\text{PBH}}{Y_B^0}\,\frac{M_P^2\,m_\nu}{v^2} &\text{for } M_1 > \Tbh^\text{in} \, ,
    \end{cases}\label{eq:m1bound1}
\end{equation}
where $Y_B^0$ denotes the observed baryon asymmetry at $T=T_0$ and $Y_\text{PBH}^\text{evap}=n_\text{PBH}/s$ at $T=\Tev$, with $n_\text{PBH}=\rho_\text{PBH}/\Min$ being the number density of PBH at evaporation~\cite{Fujita:2014hha}. Here $\zeta$ parametrizes a possible entropy production after PBH evaporation until now, i.e., $\zeta\,\left(sa^3\right)_\text{evap}=\left(sa^3\right)_0$, that can occur, for example, due to a long lived moduli field~\cite{Fujita:2014hha,Masina:2020xhk}. We will always consider $\zeta=1$, unless otherwise mentioned. To ensure non-thermal production of baryon asymmetry it is also necessary to consider $M_1 > T_\text{evap}$~\cite{Fujita:2014hha} that leads to
\begin{equation}
M_1 \gtrsim 3\times 10^{-3}\,\left[\mathcal{G}^2\,g_\star(\Tev)\left(\frac{M_P^5}{\Min^3}\right)^2\right]^{1/4}.
\label{eq:m1bound2}
\end{equation}
Otherwise, for $M_1<T_\text{evap}$, the RHNs produced from PBH evaporation are in thermal bath and then washout processes are in effect. Finally, in order for lepton asymmetry to be sufficiently generated from RHNs produced from PBH evaporation, one requires evaporation to be over before sphaleron decoupling temperature $\Tev\gtrsim T_\text{EW}$, which translates into an upper bound on initial PBH mass $\Min\lesssim 3\times 10^5\,\text{g}$. 
\begin{figure}[htb!]
    \centering
    \includegraphics[scale=0.42]{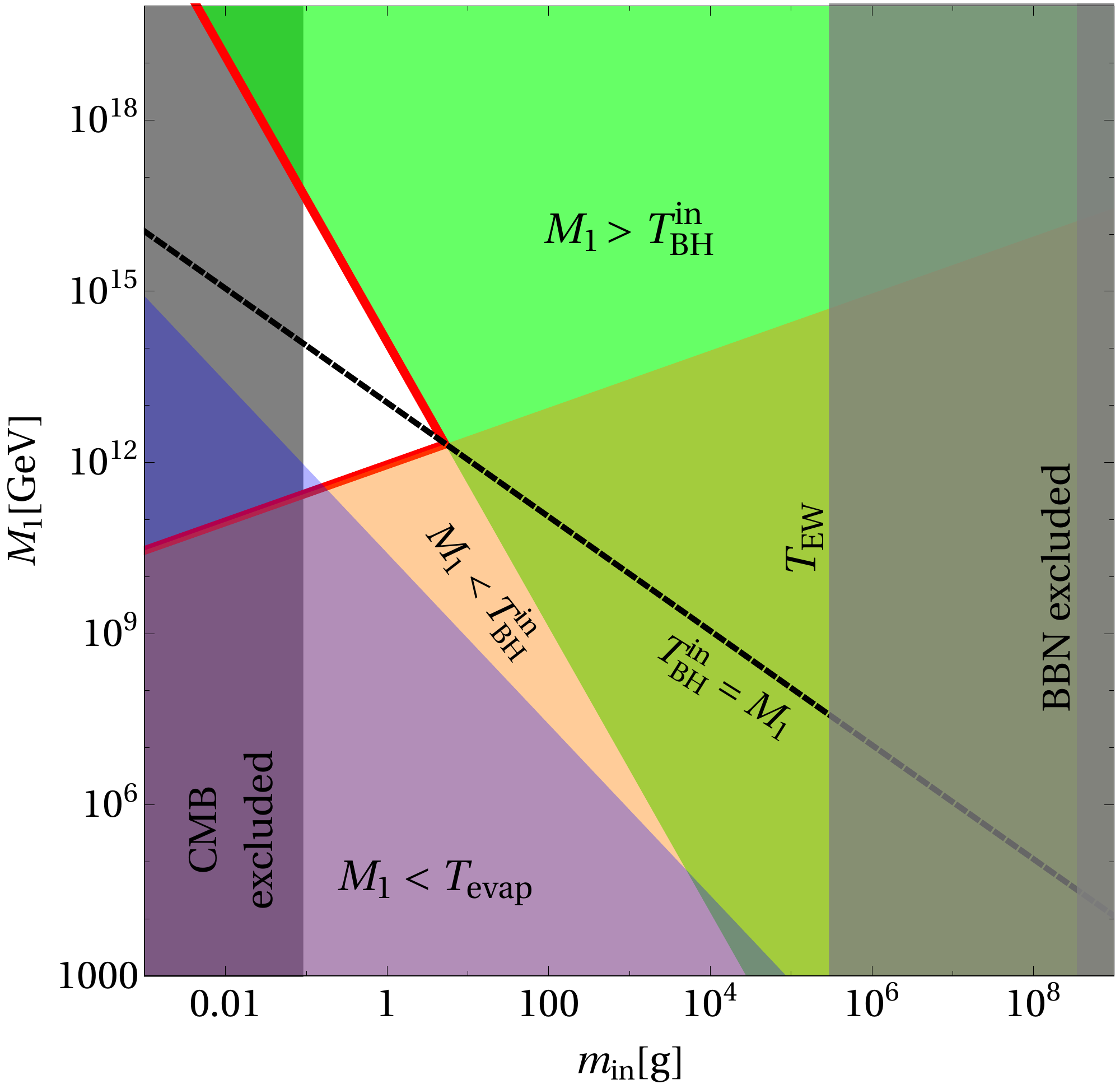}
    \caption{Bound on RHN mass from the requirement of obtaining observed baryon asymmetry from PBH evaporation. All the shaded regions are discarded, while the red solid line segregates the two regions corresponding to the bounds derived in Eq.~\eqref{eq:m1bound1} and Eq.~\eqref{eq:m1bound2}. The white triangular region in the middle is the region that is allowed.}
    \label{fig:m1-bound}
\end{figure}

In Fig.~\ref{fig:m1-bound} we summarise the bounds on the lightest RHN mass, required to produce the observed baryon asymmetry, considering a purely non-thermal regime where RHN is produced from PBH evaporation. The tiny triangular white part is the only window where $Y_B=Y_B^\text{obs}$. This region typically corresponds to $0.1 \, {\rm g}\lesssim \Min\lesssim 20$ g and $10^{12}\lesssim M_1\lesssim 10^{17}$ GeV when entropy is assumed to be conserved\footnote{The viable region of baryogenesis via leptogenesis from PBH can be modified in case there is substantial entropy injection, and also considering resonance enhancement in the generation of CP asymmetry~\cite{Barman:2022gjo}. }. 

\subsection{Baryogenesis from PBH}
\label{sec:baryo}
\begin{figure}[htb!]
\centering
\includegraphics[scale=0.42]{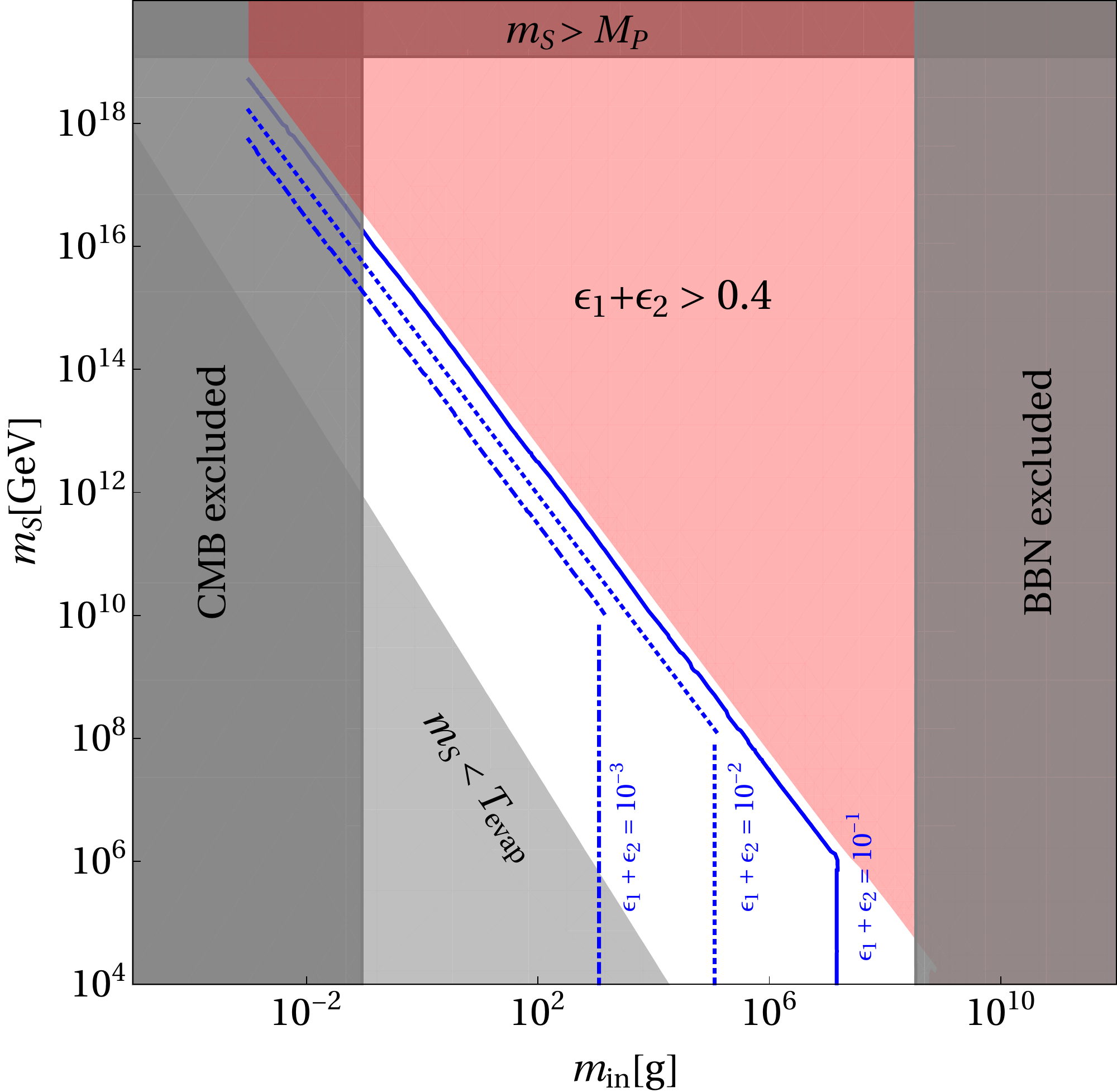}
\caption{Contours satisfying $Y_B^\text{obs}$ during PBH domination for different choices of the asymmetries, considering $\epsilon_1+\epsilon_2$. The shaded vertical regions are ruled out. The shaded lower triangular region leads to thermal baryogenesis where our analysis based on non-thermal approximations is not applicable. {{The red shaded region is where the CP asymmetry exceeds the maximum value that is allowed by model parameters (see text).}}}
    \label{fig:pbh-baryo}
\end{figure}

As advocated earlier, the present set-up allows us to have baryogenesis from the direct CP-violating decay of the coloured scalars $S_\alpha$. The final asymmetry, therefore, depends on the yield of $S$ from PBH evanescence. Then, from Eq.~\eqref{eq:etaB}, during PBH domination we obtain\footnote{As we will see, in most of our parameter space, the baryon asymmetry is created below the electroweak scale ($m_{\rm in}\gtrsim 10^{5}$ g). If the asymmetry is created above the electroweak scale, it can be washed out by sphalerons if $B-L$ number is conserved. But, since in our scenario we have $B-L$ violating interactions, such washouts would be absent.  However, there exists a conversion factor because of sphaleron which is of $\mathcal{O}(1)$ \cite{Harvey:1990qw}.}
\begin{align}\label{eq:pbh-baryo}
& Y_B(T_0) \equiv Y_B(\Tev) = \left(\epsilon_1+\epsilon_2\right)\,\frac{n_S}{s}\Bigg|_{\Tev}  = \left(\epsilon_1+\epsilon_2\right)\,\mathcal{N}_X\,\frac{n_\text{PBH}}{s}\Bigg|_{\Tev}
\nonumber\\&
= \left(\epsilon_1+\epsilon_2\right)\,
\begin{cases}
\frac{g_{X,H}}{g_{\star,H}}\,\frac{5}{\pi^2\,g_{\star s}(\Tev)}\,\left(\frac{\pi^3\,g_\star(\Tev)}{5}\right)^\frac{3}{4}\,\sqrt{\frac{\mathcal{G}\,g_{\star,H}}{10640\,\pi}}\,\sqrt{\frac{M_P}{\Min}} &\text{for } m_S < T_\text{BH}^\text{in}\,,\\[8pt]
\frac{g_{X,H}}{g_{\star,H}}\,\frac{5}{64\pi^4\,g_{\star s(\Tev)}}\,\left(\frac{\pi^3\,g_\star(\Tev)}{5}\right)^\frac{3}{4}\,\sqrt{\frac{\mathcal{G}\,g_{\star,H}}{10640\,\pi}}\,\left(\frac{M_P^9}{\Min^5\,m_S^4}\right)^\frac{1}{2} &\text{for } m_S > T_\text{BH}^\text{in}\,,
\end{cases} 
\end{align}
where we consider $m_{S_1} \approx m_{S_2}\equiv m_S$, that leads to $n_{S_1} \approx n_{S_2}\equiv n_S$ [according to Eq.~\eqref{eq:pbh-num}]. As evident from Eq.~\eqref{eq:cp-asym}, $\epsilon_{1,2}$ are functions of the two couplings $\lambda\,,\lambda'$ and the scalar mass $m_S$. In Fig.~\ref{fig:pbh-baryo} we show the parameter space satisfying the observed baryon asymmetry in $m_S-\Min$ plane assuming PBH domination $(\beta > \beta_{\rm crit})$, for different choices of $\epsilon_1+\epsilon_2$. The contours obeying observed baryon asymmetry are independent of $m_S$ when $m_S< T^{\rm in}_{\rm BH}$, while a larger $m_S$ requires lighter PBH to produce the desired asymmetry following Eq.~\eqref{eq:pbh-baryo}. {{The red shaded region corresponds to $\epsilon_1+\epsilon_2>\left(\epsilon_1+\epsilon_2\right)_\text{max}$, where the latter is $\simeq 0.4$ for our choice of parameters, as we shall discuss in later. 

Here we would like to mention that, the condition $m_S<\Tev$ is enough to ensure the non-thermal production of asymmetry, even though in this case the new scalar carries non-trivial colour charge. In order to verify that we first note, the dominant 2-to-2 interaction of the scalar with the thermal bath takes place through $SS\to\tilde g\tilde g$ at the end of PBH evaporation, where $\tilde g$ are the gluons. The corresponding cross-section reads
\begin{align}
&\sigma (s)_{SS\to\tilde g\tilde g} = \frac{g_s^4}{576\,\pi\,s}\,\sqrt{\frac{s}{s-4\,m_S^2}}\,,    
\end{align}
where the presence of the strong coupling constant $g_s$ shows this is an irreducible process. The thermally averaged cross-section and decay rate of $S$ is given by
\begin{align}\label{eq:sigv}
& \langle\sigma v\rangle_{SS\to\tilde g\tilde g} = \frac{1}{8\,m_S^4\,T_\text{BH}\,K_2\left(m_S/T_\text{BH}\right)^2}\,\int_{4\,m_S^2}^\infty\,ds\,\sigma(s)_{SS\to\tilde g\tilde g}\,\sqrt{s}\,K_1\left(\sqrt{s}/T_\text{BH}\right)
\nonumber\\&
\langle\Gamma_S\rangle\approx\frac{K_1\left(m_S/T_\text{BH}\right)}{K_2\left(m_S/T_\text{BH}\right)}\,\Gamma_S\,,
\end{align}
where $\Gamma_S$ is given by Eq.~\eqref{eq:S-decay}. Note that, we have calculated the thermal average with respect to the PBH temperature and not the bath temperature. We find, for all PBH masses of our interest, the scattering rate $n_S\times\langle\sigma v\rangle$ stays way below the decay rate at $T=\Tev$, where 
\begin{align}
& n_S = s(\Tev)\times
\begin{cases}
\frac{g_{X,H}}{g_{\star,H}}\,\frac{5}{\pi^2\,g_{\star s}(\Tev)}\,\left(\frac{\pi^3\,g_\star(\Tev)}{5}\right)^\frac{3}{4}\,\sqrt{\frac{\mathcal{G}\,g_{\star,H}}{10640\,\pi}}\,\sqrt{\frac{M_P}{\Min}} &\text{for } m_S < T_\text{BH}^\text{in}\,,\\[8pt]
\frac{g_{X,H}}{g_{\star,H}}\,\frac{5}{64\pi^4\,g_{\star s(\Tev)}}\,\left(\frac{\pi^3\,g_\star(\Tev)}{5}\right)^\frac{3}{4}\,\sqrt{\frac{\mathcal{G}\,g_{\star,H}}{10640\,\pi}}\,\left(\frac{M_P^9}{\Min^5\,m_S^4}\right)^\frac{1}{2} &\text{for } m_S > T_\text{BH}^\text{in}\,.
\end{cases} 
\end{align}
This implies, the CP-violating decay rate of the scalar is much more efficient than its 2-to-2 scattering rate at $T=\Tev$\footnote{Following the prescription in~\cite{Cheek:2021cfe, Barman:2021ost}, we have also checked the scattering efficiency of a PBH generated $S$ scattering on a thermal $S$, and found that this rate is much below the Hubble rate for the range of PBH masses that produce right DM and baryon abundance.}. This is typically because of the suppression from $m_S$ and $\Tbh$ in Eq.~\eqref{eq:sigv}.
}}
The noteworthy point here is that a direct baryogenesis from PBH opens up a whole new PBH mass range compared to baryogenesis via leptogenesis (as seen in Fig.~\ref{fig:m1-bound}). 

\begin{figure}[htb!]
\centering
\includegraphics[scale=0.3]{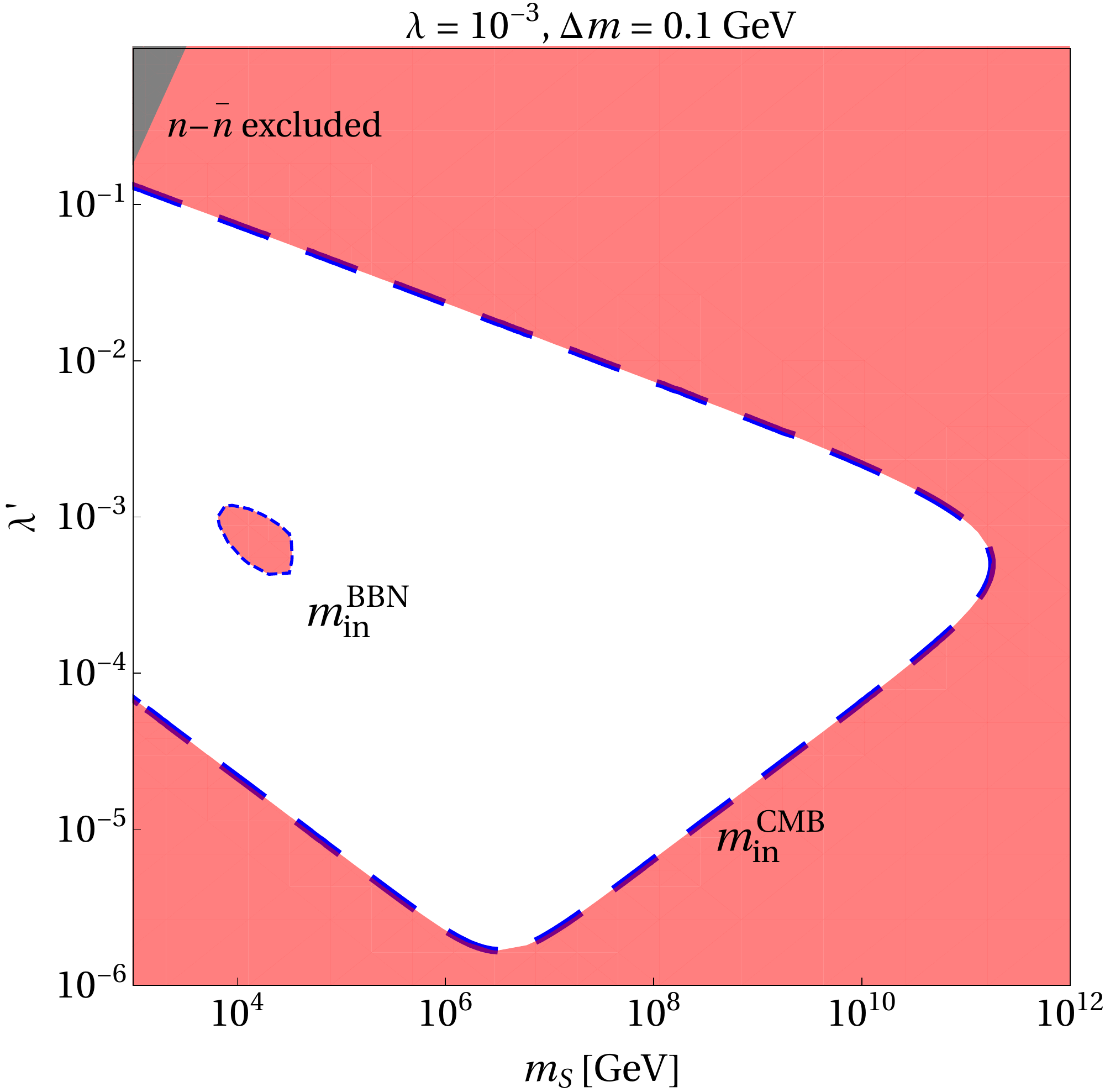}
~~~~\includegraphics[scale=0.42]{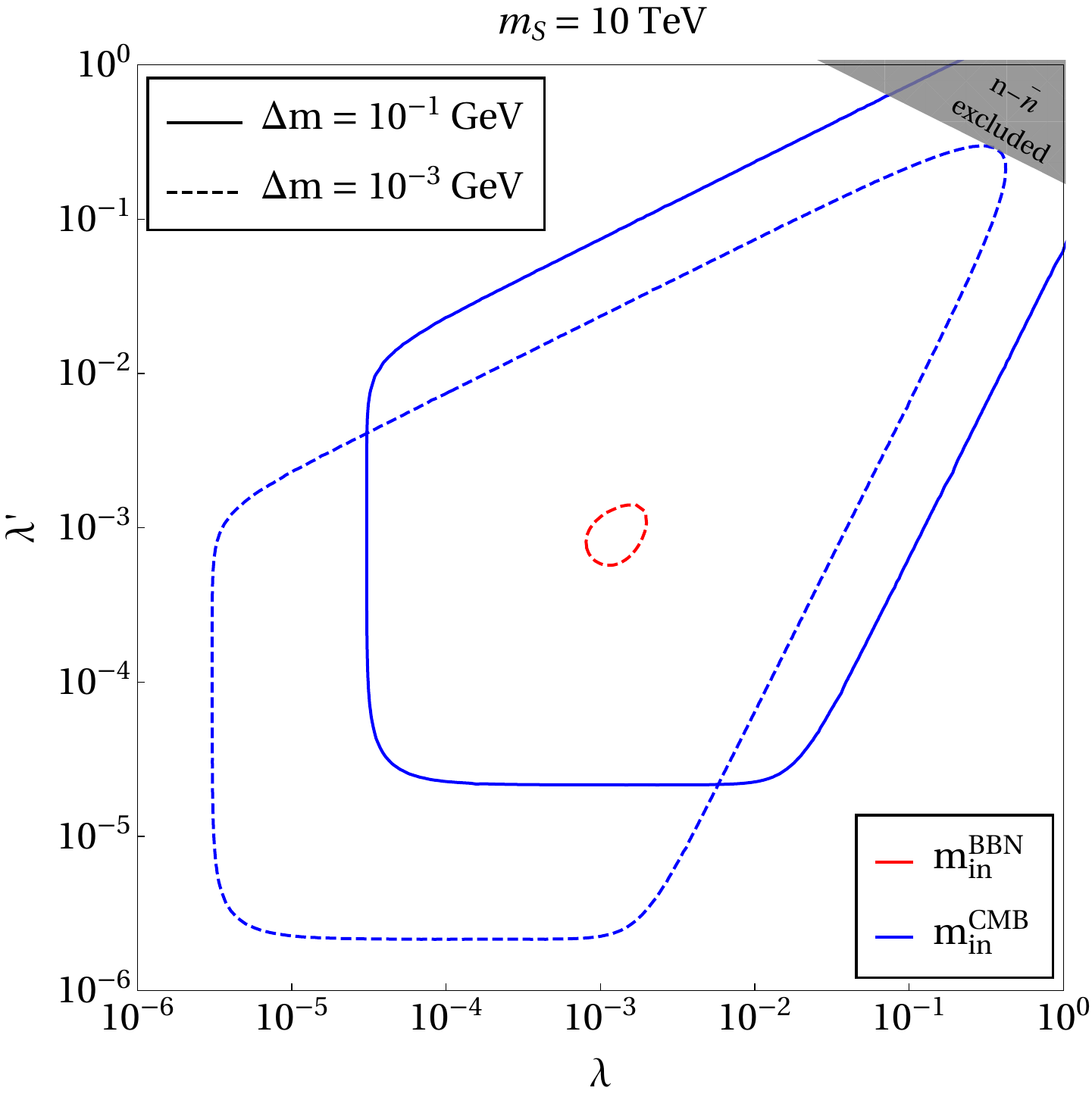}\\[10 pt]
\includegraphics[scale=0.42]{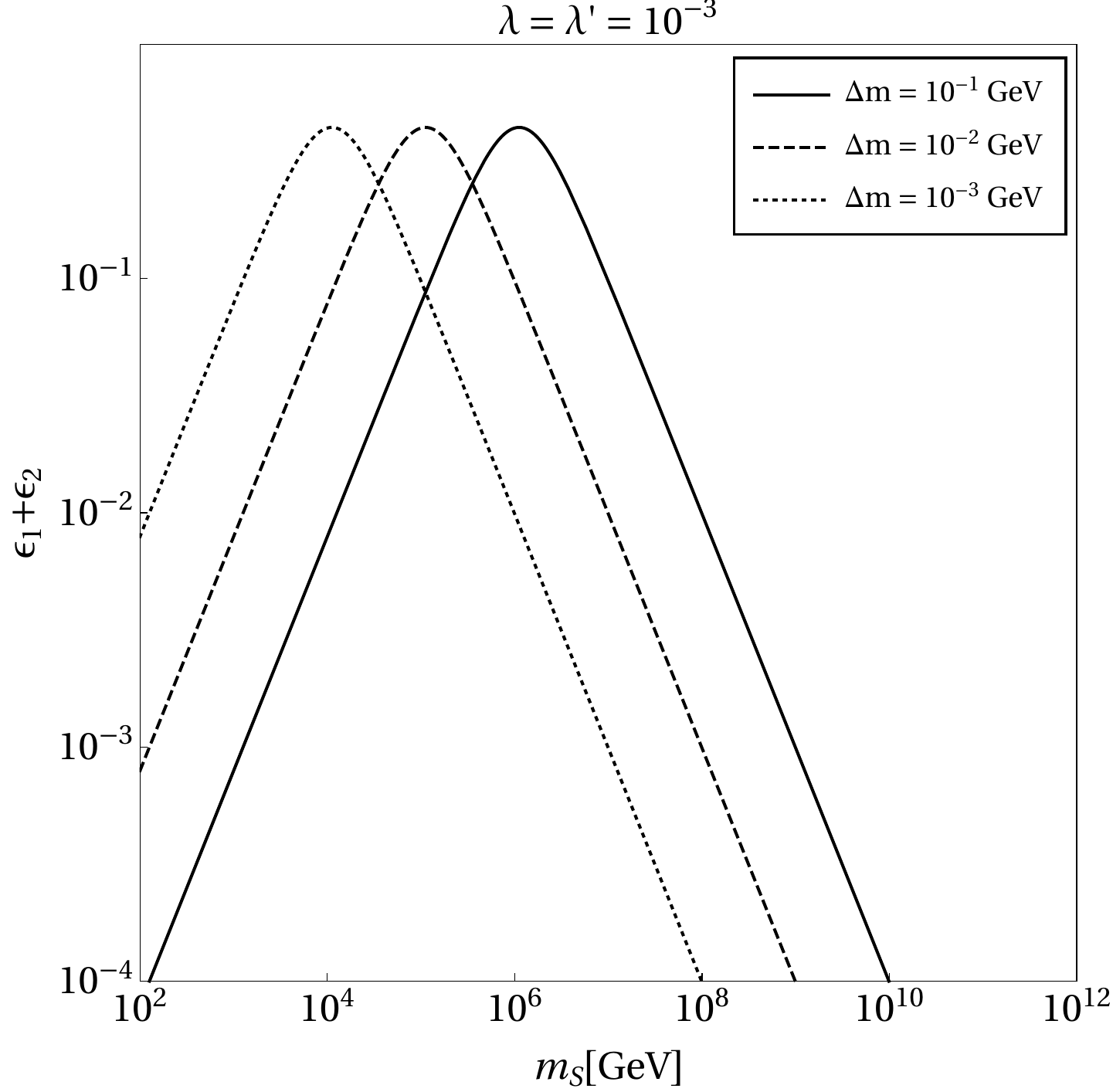}
    \caption{{\it Top Left:} Contours satisfying $Y_B^\text{obs}$ in $\lambda'-m_S$ plane for $\Min=\Min^\text{CMB}$ (blue) and $\Min=\Min^\text{BBN}$ (red). {bf{The shaded region is disallowed from BBN and CMB constraints on PBH mass (see text)}}. {\it Top Right:} Contours of $Y_B^\text{obs}$ in $\lambda'-\lambda$ plane for two different choices of $\Delta m$ shown in solid and dashed lines. The blue and red contours correspond to $\Min=\Min^\text{CMB}$ and $\Min=\Min^\text{BBN}$ respectively. {\it Bottom:} Variation of CP asymmetry as a function of $m_S$. In all cases we choose $m_\psi=1$ GeV. The shaded regions are disfavoured from $n-\bar{n}$ oscillation limits [cf.Eq.~\eqref{eq:nnbar}].}
    \label{fig:yb-cntr}
\end{figure}
 
Possible range of coupling values that can give rise to the observed BAU, are shown in Fig.~\ref{fig:yb-cntr}. In the {\it top left} panel we plot contours of correct baryon asymmetry in $\lambda'-m_S$ plane for a fixed $\Delta m$, and choosing the minimum and maximum PBH masses that satisfy CMB and BBN bound respectively. The blue coloured contour corresponds to $\Min=\Min^\text{CMB}\sim 0.1$ g, while the red coloured contour is for $\Min=\Min^\text{BBN}\sim 10^8$ g. Thus, the region in between these two can produce $Y_B^\text{obs}$ for gradually decreasing PBH mass, as we move from the red to the blue contour. {{The red shaded regions are disallowed as they require PBH lighter (heavier) than the one viable from CMB (BBN) bound}}. Note that, because of $\Delta m\ll m_S$, we do not discriminate between two scalar masses and refer to them as $m_S$. Note that, large Yukawa or lighter $m_S$ are forbidden (shown by the grey shaded region) as they give rise to $n-\bar{n}$ oscillations with small lifetime, in conflict with the bound on Eq.~\eqref{eq:nnbar}. In the {\it top right} panel we project the allowed parameter space in $\lambda'-\lambda$ plane for a fixed $m_S=10$ TeV, and for different choices of $\Delta m$, shown by blue coloured solid and dashed contours. The red and blue coloured contours are for PBH of masses allowed from BBN and from CMB bounds respectively. {{The region outside each blue contour is discarded, as they correspond to PBH masses lighter than the one allowed from CMB measurements, as shown by the red shaded region in the top left panel. However, as these contours overlap on each other, depending on the choice of $\Delta m$, we refrain from shading those regions}}. For each case we see that the correct asymmetry can be found twice as the CP-asymmetry shows a resonance behaviour, which becomes maximum when $\Delta m\sim\Gamma_S/2$, as shown in the {\it bottom} panel. This results in closed contours for right baryon asymmetry. Clearly, if we deviate from the resonance regime, the required CP asymmetry can be obtained by suitable choices of Yukawa couplings. {{We find, a maximum of CP asymmetry of $\left(\epsilon_1+\epsilon_2\right)_\text{max}=0.43$ is possible for parameters of our interest. We thus discard values larger than $\left(\epsilon_1+\epsilon_2\right)_\text{max}$, as shown by the red shaded region in Fig.~\ref{fig:pbh-baryo}}}. As the {\it top right} panel figure shows, increasing the mass splitting $\Delta m$ corresponds to larger Yukawa couplings in order to obtain the desired asymmetry. {{Here we would like to remind the readers once again that the results so far apply for PBH dominated epoch, i.e., $\beta>\beta_\text{crit}$.}}

As noted earlier, the simple estimate for baryon asymmetry adopted here holds in the non-thermal ballpark only. While PBH evaporation at late epoch $(T_{\rm evap} < m_S)$ leads to such non-thermal production of $S$, it is possible to have thermally generated $S$ too given the fact that prior to PBH domination, the universe is assumed to be radiation dominated. However, being a coloured scalar, $S$ will be in equilibrium for a long epoch due to strong annihilations into gluons (independent of the Yukawas $\lambda\,,\lambda'$) leading to dilution in generated asymmetry. On the other hand, the out of equilibrium criteria of its decay will force the corresponding Yukawa couplings $(\lambda, \lambda^\prime)$ to be in smaller regime, reducing the CP-asymmetry~\cite{Allahverdi:2013mza, Allahverdi:2017edd}. In other low scale baryogenesis scenarios like~\cite{Claudson:1983js} and post-sphaleron baryogenesis~\cite{Babu:2006xc}, this is ensured by considering the decaying particle to be colour neutral whose baryon number violating decay into quarks create the asymmetry without requiring any non-thermal origin. Additionally, there will be washouts from inverse decay and scattering further reducing the baryon asymmetry produced from thermally generated $S$. Finally, PBH evaporation at late epoch is likely to cause entropy dilution to any baryon asymmetry generated at higher temperatures $(T> T_{\rm evap})$, as noticed in PBH-generated leptogenesis works earlier \cite{Perez-Gonzalez:2020vnz, JyotiDas:2021shi}. Therefore, the thermally generated baryon asymmetry can be ignored in our setup validating the non-thermal estimates discussed above.

\subsection{A common parameter space for baryogenesis and dark matter}
\label{sec:dm}
\begin{figure}[htb!]
\centering
\includegraphics[scale=0.35]{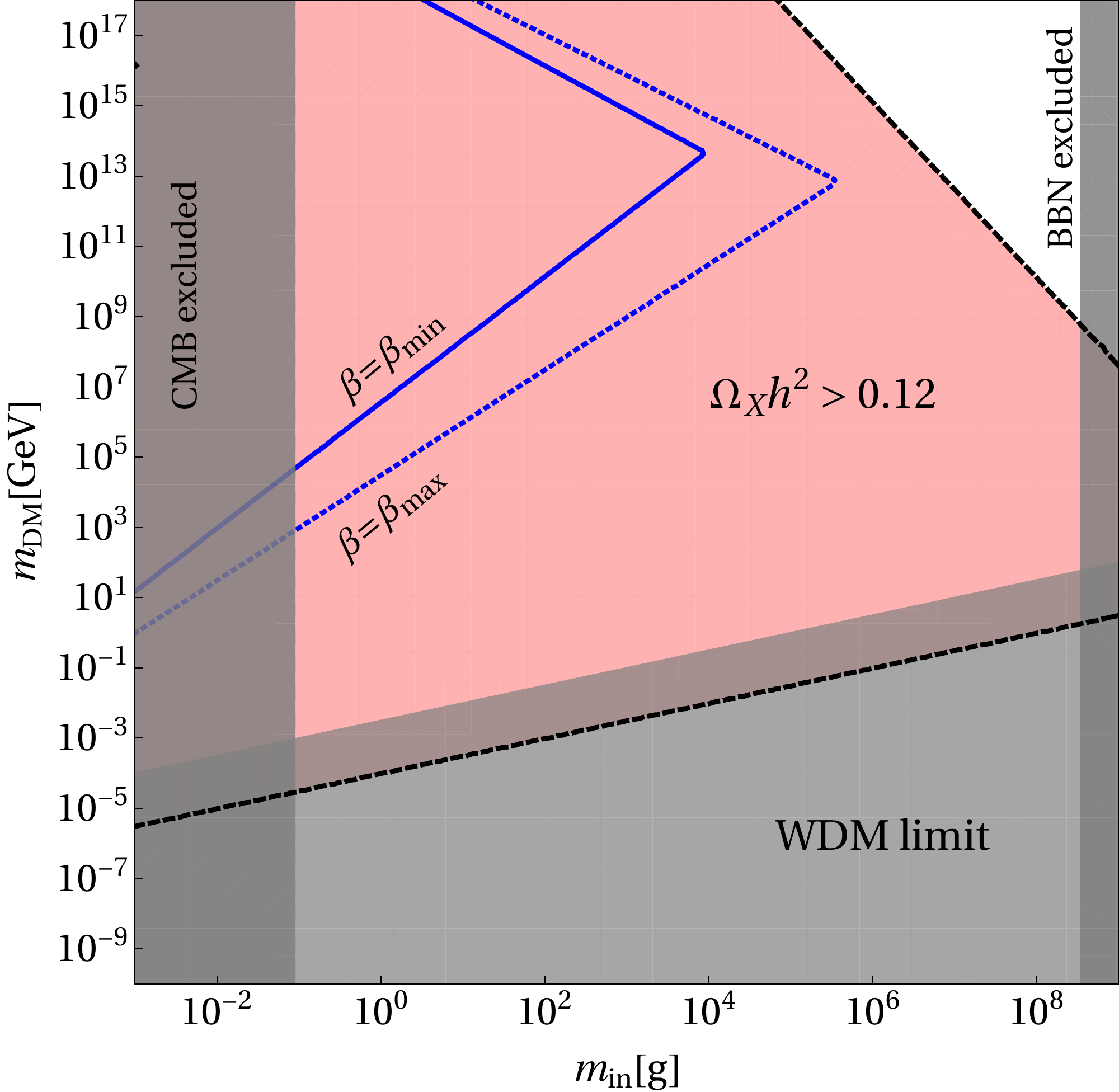}~~~~\includegraphics[scale=0.47]{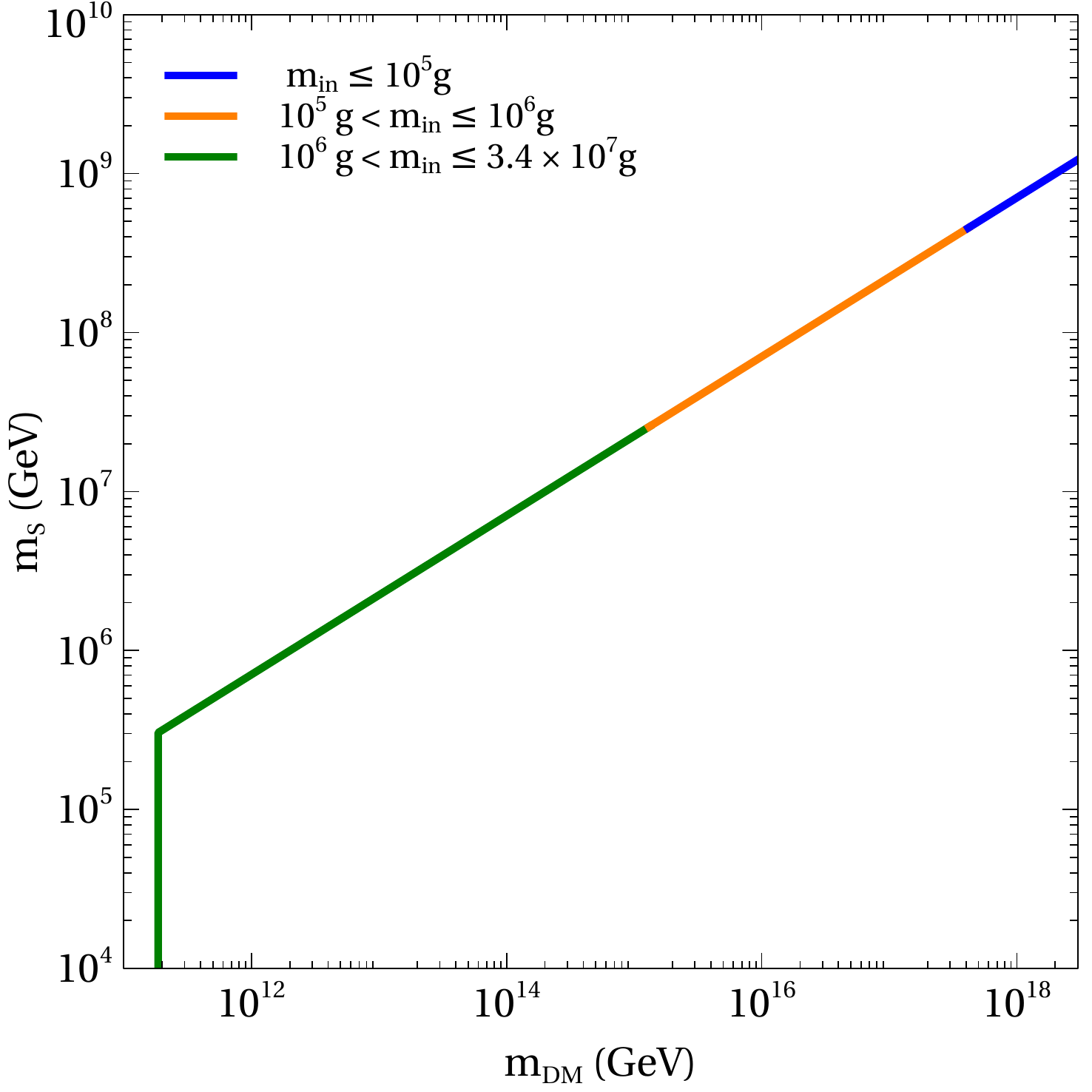}
    \caption{{\it Left:} The red coloured region of over-abundance for DM produced entirely from PBH evaporation. We also show the warm DM constraint, effective in the region of lower mass DM. Along the blue contours right relic abundance is produced considering gravitational UV freeze-in for lower (solid) and upper (broken) bound on $\beta$ [cf. Eq.~\eqref{eq:beta}]. {{Along the black dashed contour right abundance is obtained considering the entire DM is produced from PBH evaporation}}. {\it Right:} Viable parameter space satisfying relic density and baryon asymmetry in $m_\text{DM}-m_S$ plane for a fixed $\epsilon_1+\epsilon_2=0.1$ and scanned over a range of $\Min$, shown with different colours. {{In all cases we consider $\beta>\beta_\text{crit}$.}}}
    \label{fig:dm}
\end{figure}

DM of arbitrary intrinsic spin can be produced directly from PBH evaporation. Thus, the DM abundance can be expressed as
\begin{align}\label{eq:dm-yld}
& Y_\text{DM}(T_0)\equiv\frac{n_\text{DM}}{s}\Bigg|_{T_0}=  \frac{3}{4}\,\frac{\gsH(\Tin)}{\gss}\,N_\text{DM}\times
\begin{cases}
\beta\frac{\Tin}{\Min}\,,\beta\leq\beta_c \\
\frac{\bar{T}_\text{evap}}{\Min}\,,\beta\geq\beta_c\,,
\end{cases}
\end{align}
where $N_\text{DM}\equiv\mathcal{N}_X$ as given in Eq.~\eqref{eq:pbh-num}. Thus, the DM relic abundance $\Omega_\text{DM}\,h^2 = \frac{m_\text{DM}\,s_0}{\rho_c}\,Y_\text{DM}\left(T_0\right)\,,$ at the present epoch reads 
\begin{equation}
    \Omega_\text{DM}\,h^2 =\mathbb{C}(T_\text{ev})
    \begin{cases}
      \frac{1}{\pi^2}\,\sqrt{\frac{M_P}{m_\text{in}}}\,m_\text{DM} &\text{for } m_\text{DM} < T^{\rm in}_{\rm BH}\,,\\[8pt]
       \frac{1}{64\,\pi^4}\left(\frac{M_P}{m_\text{in}}\right)^{5/2}\,\frac{M_P^2}{m_\text{DM}} &\text{for } m_\text{DM} > T^{\rm in}_{\rm BH}\,,
    \end{cases}\label{eq:rel-dm}\,
\end{equation}
with $\mathbb{C}(T_\text{ev})=\frac{s_0}{\rho_c}\,\frac{1}{\zeta}\,\frac{g_{X,H}}{g_{\star,H}}\,\frac{5}{g_{\star s}(T_\text{evap})}\,\,\left(\frac{\pi^3\,g_{\star}(T_\text{evap})}{5}\right)^{3/4}\,\sqrt{\frac{\mathcal{G}\,g_{\star,H}}{10640\,\pi}}$. {{Note that, Eq.~\eqref{eq:rel-dm}} is valid for PBH dominated epoch}. It is worth mentioning that if we assume a PBH dominated era, then for majority of the parameter space the DM gets over-produced from PBH evaporation irrespective of their spins and only $m_{\rm DM} \gtrsim 10^{10}$ GeV can lead to right abundance for $\Min\gtrsim 10^6$ g~\cite{Fujita:2014hha, Samanta:2021mdm}. DM over-production can also be controlled by choosing sufficiently light DM mass, but it is likely to face constraints from structure formation. In order not to spoil the structure formation, a DM candidate which is part of the thermal bath or produced from the thermal bath should have mass above a few keV (depending upon the details of the production mechanism) in order to give required free-streaming length of DM as constrained from Lyman-$\alpha$ flux-power spectra~\cite{Irsic:2017ixq, Ballesteros:2020adh, DEramo:2020gpr}. Such light DM of keV scale leads to a warm dark matter (WDM) scenario having free-streaming length in the intermediate range relative to that of cold and hot DM. If such light DM is also produced from PBH evaporation, it leads to a potential hot component in total DM abundance, tightly constrained by observations related to the CMB and baryon acoustic oscillation (BAO) leading to an upper bound on the fraction of this hot component with respect to the total DM, depending on the value of DM mass~\cite{Diamanti:2017xfo}. The lower bound on DM mass from Lyman-$\alpha$ can be found in~\cite{Fujita:2014hha,Masina:2020xhk,Barman:2022gjo}. As for the heavy DM case relic abundance has an inverse dependence on the DM mass, hence heavy DM leads to under-abundance. This is shown in the left panel Fig.~\ref{fig:dm}, considering the DM to be of spin zero. {{The white region in the top right corner is where the DM is under abundant, while along the black dashed line right relic abundance is obtained, without considering any extra source of entropy injection after PBH evaporation}}. Now, for the simultaneous realization of both the observed DM relic density and the baryon asymmetry, $m_S$ and $m_{\text{DM}}$ become connected through Eq. \eqref{eq:pbh-baryo} and \eqref{eq:rel-dm}. In the right panel of Fig.~\ref{fig:dm}, we show the contour satisfying both DM relic density and baryon asymmetry, considering $\epsilon_1+\epsilon_2=0.1$. The range of $m_{\rm in}$ which varies along this contour is shown with different colours. For $m_{\rm in}\lesssim 2.5 \times 10^{4}$ g, DM mass becomes super-Planckian. On the other hand, for $m_{\rm in}\gtrsim 3.4 \times 10^{7}$ g, $m_S$ becomes less than $T_{\rm BH}^{\rm in}$ and hence $Y_B$ becomes independent of $m_S$ (cf. Eq.~\eqref{eq:pbh-baryo}). {{Note that,the parameter space features same characteristic as in Fig.~\ref{fig:pbh-baryo} because of the requirement of simultaneously satisfying right relic abundance and baryon asymmetry.}}

Apart from PBH, pure gravitational production of DM can also take place from the 2-to-2 scattering of the bath particles via $s$-channel mediation of massless graviton. The interaction rate density for such a process reads~\cite{Garny:2015sjg, Tang:2017hvq,Garny:2017kha, Bernal:2018qlk,Barman:2021ugy,Barman:2021qds}
\begin{equation}
    \gamma(T) = \alpha\, \frac{T^8}{M_P^4}\,,
\end{equation}
with $\alpha \simeq 1.9\times 10^{-4}$ (real scalar), $\alpha \simeq 1.1\times 10^{-3}$ (Dirac fermion) or $\alpha \simeq 2.3\times 10^{-3}$ (vector boson). This kind of production is unavoidable due to universal coupling between the gravity and the stress-energy tensor involving the matter particles. The BEQ governing the time evolution of DM number density is thus given by
\begin{equation}\label{eq:beq-UV}
\dot n_\text{DM} + 3\,H\,n_\text{DM}=\gamma\,.    
\end{equation}
For temperatures much lower than the reheat temperature i.e., $T\ll\Trh$, the DM yield can be analytically obtained by integrating Eq.~\eqref{eq:beq-UV}
\begin{equation}\label{eq:grav-yield1}
Y_0 = \frac{45\,\alpha}{2\,\pi^3\,g_{\star s}}\,\sqrt{\frac{10}{g_\star}}\,\left(\frac{T_\text{rh}}{M_P}\right)^3\,,    
\end{equation}
where we define the DM yield as $Y\equiv n_\text{DM}/s$, with $s=\frac{2\,\pi^2}{45}\,g_{\star s}\,T^3$ and consider $m_\text{DM}\ll\Trh$. On the other hand, if the DM mass is such that $\Trh\ll m_\text{DM}\ll \Tmax$, where $\Tmax$ corresponds to the maximum temperature during reheating, then the DM can be produced during but not after the reheating. In the case the DM yield can be obtained by integrating Eq.~\eqref{eq:beq-UV} for $\Tmax\geq m_\text{DM}\geq\Trh$
\begin{equation}\label{eq:grav-yield2}
Y_0 =  \frac{45\,\alpha}{2\,\pi^3\,g_{\star s}}\,\sqrt{\frac{10}{g_\star}}\,\frac{\Trh^7}{M_P^3\,m_\text{DM}^4}\,.     
\end{equation}
Now, the DM produced via gravitational UV freeze-in shall undergo dilution due to evaporation of the PBH, that can be quantified as~\cite{Bernal:2021yyb,Bernal:2022oha} 
\begin{align}
& \frac{S(T_\text{in})}{S(T_\text{evap})}\simeq
\begin{cases}
1\,,~\beta < \beta_c\\[10pt]
\frac{T_\text{evap}}{T_\text{peq}}\simeq 10^{-2} \left(\frac{M_P}{\Min}\right)^\frac{3}{2}\,\frac{M_P}{\beta\,T_\text{in}}\,,~\beta>\beta_c\,,
\end{cases}
\end{align}
where we define $S=a^3\,s(T)$. The temperature $T_\text{peq}$ is defined as the epoch of equality between SM radiation and the PBH energy densities $\rho_R(T_\text{peq})=\rho_\text{BH}(T_\text{peq})$, and is given by
\begin{equation}
T_\text{peq} = \beta\,T_\text{in}\,\left(\frac{g_{\star,s}(T_\text{in})}{g_{\star,s}(T_\text{in})}\right)^\frac{1}{3}\,.    
\end{equation}
The observed DM abundance can then be achieved 
\begin{equation}
m_\text{DM}\,Y_0\,\frac{S(T_\text{in})}{S(T_\text{evap})}=\Omega_\text{DM}\,h^2\,\frac{1}{s_0}\,\frac{\rho_c}{h^2}\simeq 4.3\times 10^{-10}\,\rm GeV\,,   
\end{equation}
with $\rho_c$ being the critical density of the universe. Along the blue contours in the left panel of Fig.~\ref{fig:dm}, correct DM abundance via gravitational UV freeze-in is obtained considering $\Trh=T_\text{in}$ and a spin-0 DM candidate. The two contours, shown via solid and dotted curves correspond to the lower and upper bound on $\beta$, following Eq.~\eqref{eq:beta}. To the left of each contours, DM is over-produced due to gravitational UV freeze-in. This shows, the region of over-abundance corresponding to gravitational UV freeze-in overlaps with the region of over-abundance for DM production from PBH evaporation. 

It should be noted that, we have considered a spin zero scalar to be the DM candidate in the above discussion and shown it to satisfy the correct relic abundance criteria, together with correct BAU produced from PBH evaporation, for superheavy DM masses. While the original model has a DM candidate $\psi$ provided its mass lies in the tiny window $m_p-m_e \leq m_\psi \leq m_p+m_e$ and its coupling to SM leptons and Higgs are forbidden by some additional symmetries. However, as the above discussion shows, such DM with mass around 1 GeV is likely to be over-produced for the PBH mass range and initial fraction $(\beta > \beta_{\rm crit})$ we are choosing to have desired baryogenesis and GW spectrum. In earlier works, for example \cite{Allahverdi:2017edd}, such light DM was found to be thermally over-produced requiring the non-thermal production from moduli decay. While non-thermal production of DM from a moduli dominated era can be controlled by suitable choice of couplings, this freedom no longer exists in a PBH dominated era due to democratic gravitational couplings to all particles. Therefore, we do not need to choose $\psi$ mass in the above-mentioned tiny window, it can be mass larger than proton mass and can also play some role in generating light neutrino masses via its coupling to SM leptons and Higgs, in a way similar to \cite{Dev:2015uca,Davoudiasl:2015jja}.

\section{Induced Gravitational Waves from PBH Density Fluctuations}
\label{sec:GW}
There are several ways in which PBHs can be involved in the production of primordial gravitational waves. First, the large curvature perturbations that form the PBH can induce GWs \cite{Saito:2008jc}. Secondly, PBH can radiate gravitons through Hawking evaporation which leads to high-frequency GWs \cite{Anantua:2008am}. Next, PBHs can form mergers which can also emit GWs \cite{Zagorac:2019ekv}. Finally, the inhomogeneous distribution of PBHs also leads to density fluctuations that can induce GWs \cite{Papanikolaou:2020qtd, Domenech:2020ssp, Inomata:2020lmk}. In the present work, we focus on this last scenario. 
\begin{figure}[htb!]
\centering
\includegraphics[scale=0.4]{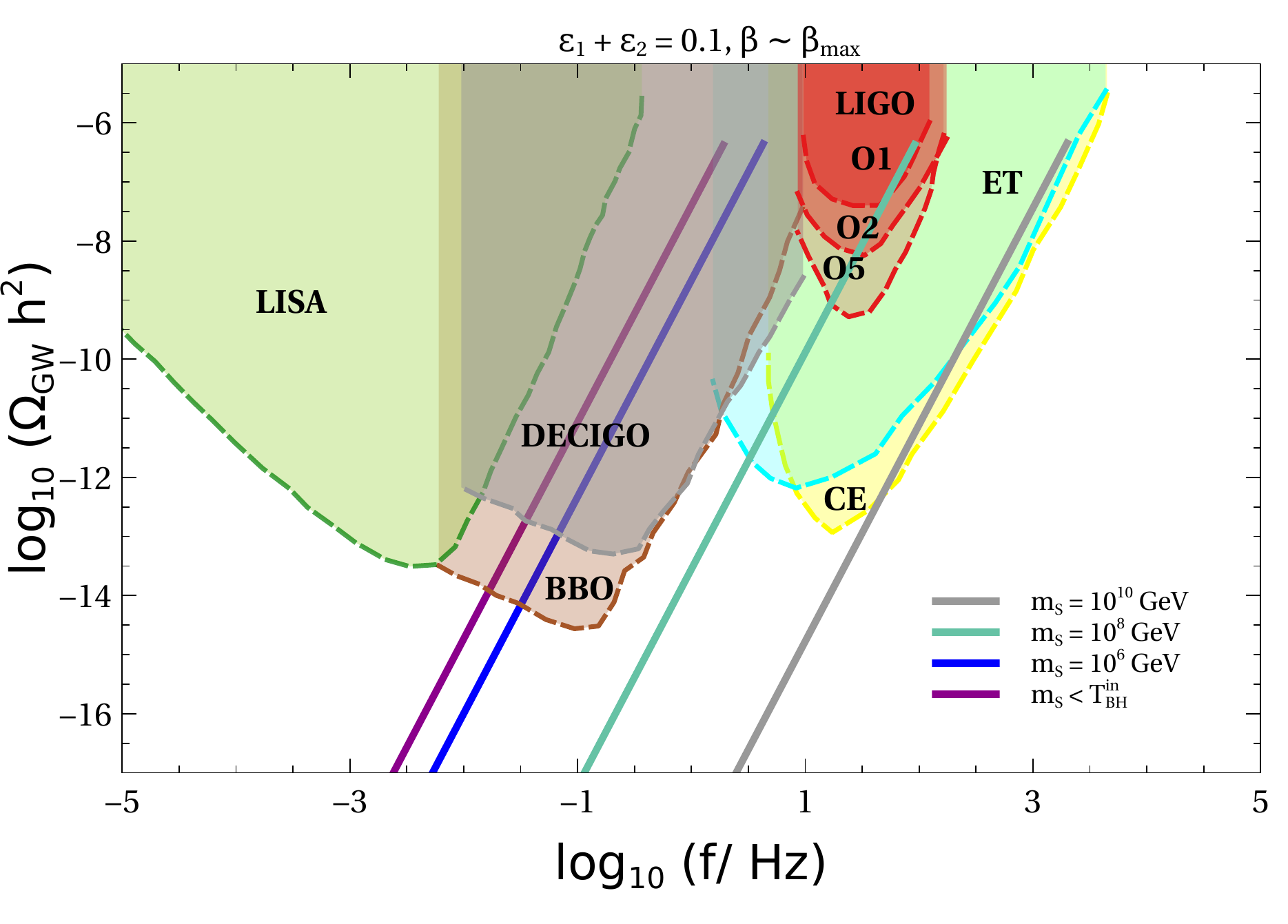}
 ~~~~\includegraphics[scale=0.4]{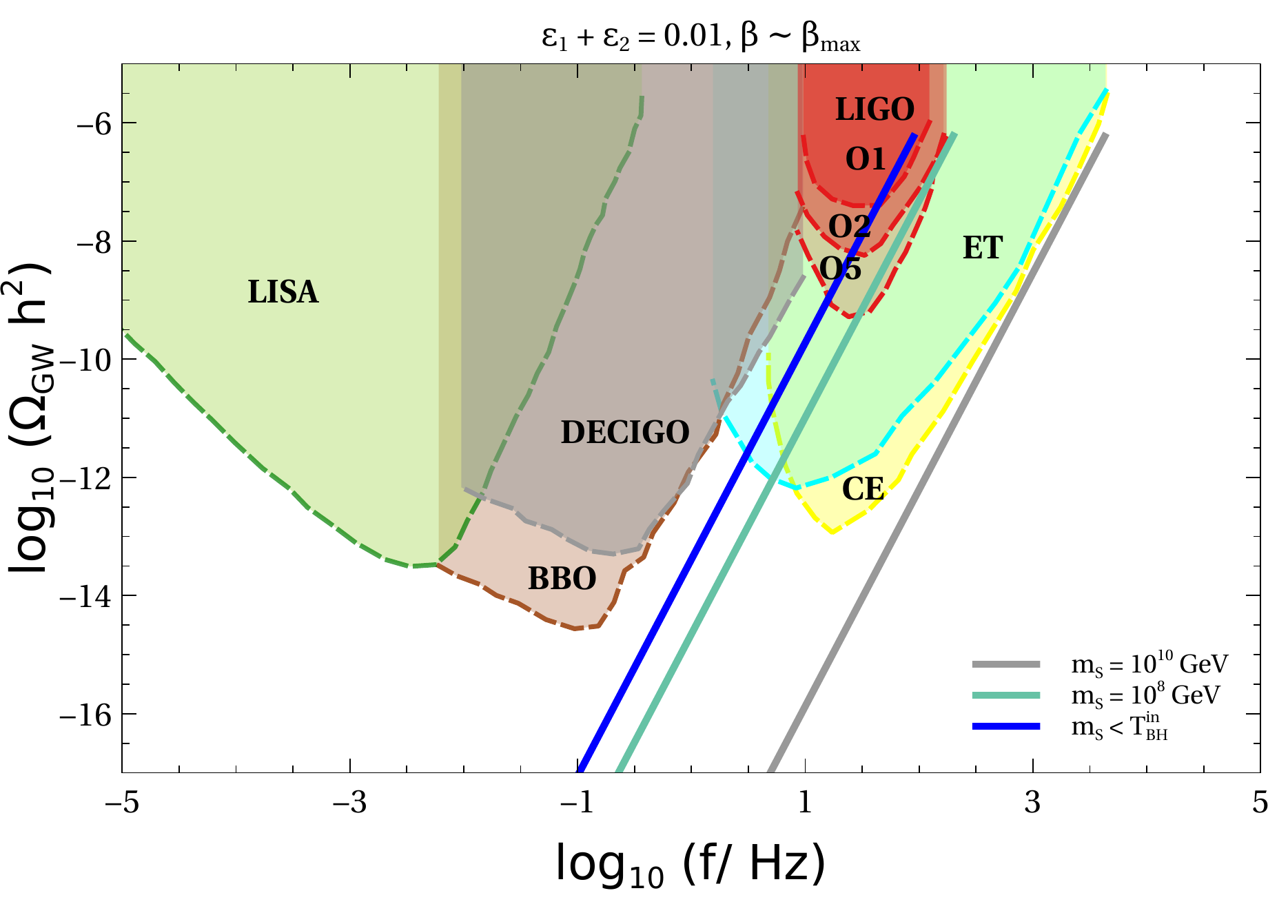}
    \caption{GW spectra induced from PBH density fluctuations, from the requirement of obtaining the observed baryon asymmetry $Y_{B}^{\text{obs}}$, with $\epsilon_1+\epsilon_2=0.1$ (left panel) and $\epsilon_1+\epsilon_2=0.01$ (right panel). The experimental sensitivities of BBO~\cite{Crowder:2005nr,Corbin:2005ny,Harry:2006fi}, DECIGO~\cite{Seto:2001qf,Kawamura:2006up,Yagi:2011wg}, CE~\cite{LIGOScientific:2016wof,Reitze:2019iox}, ET~\cite{Punturo:2010zz, Hild:2010id,Sathyaprakash:2012jk, Maggiore:2019uih}, LISA~\cite{2017arXiv170200786A} and aLIGO/VIRGO~\cite{LIGOScientific:2014qfs, LIGOScientific:2016wof, LIGOScientific:2016jlg}. Here we use the sensitivity curves derived in Ref.~\cite{Schmitz:2020syl} are shown as shaded regions of different colours.}
    \label{fig:pbh-igw}
\end{figure}
\begin{figure}[htb!]
\centering
\includegraphics[scale=0.45]{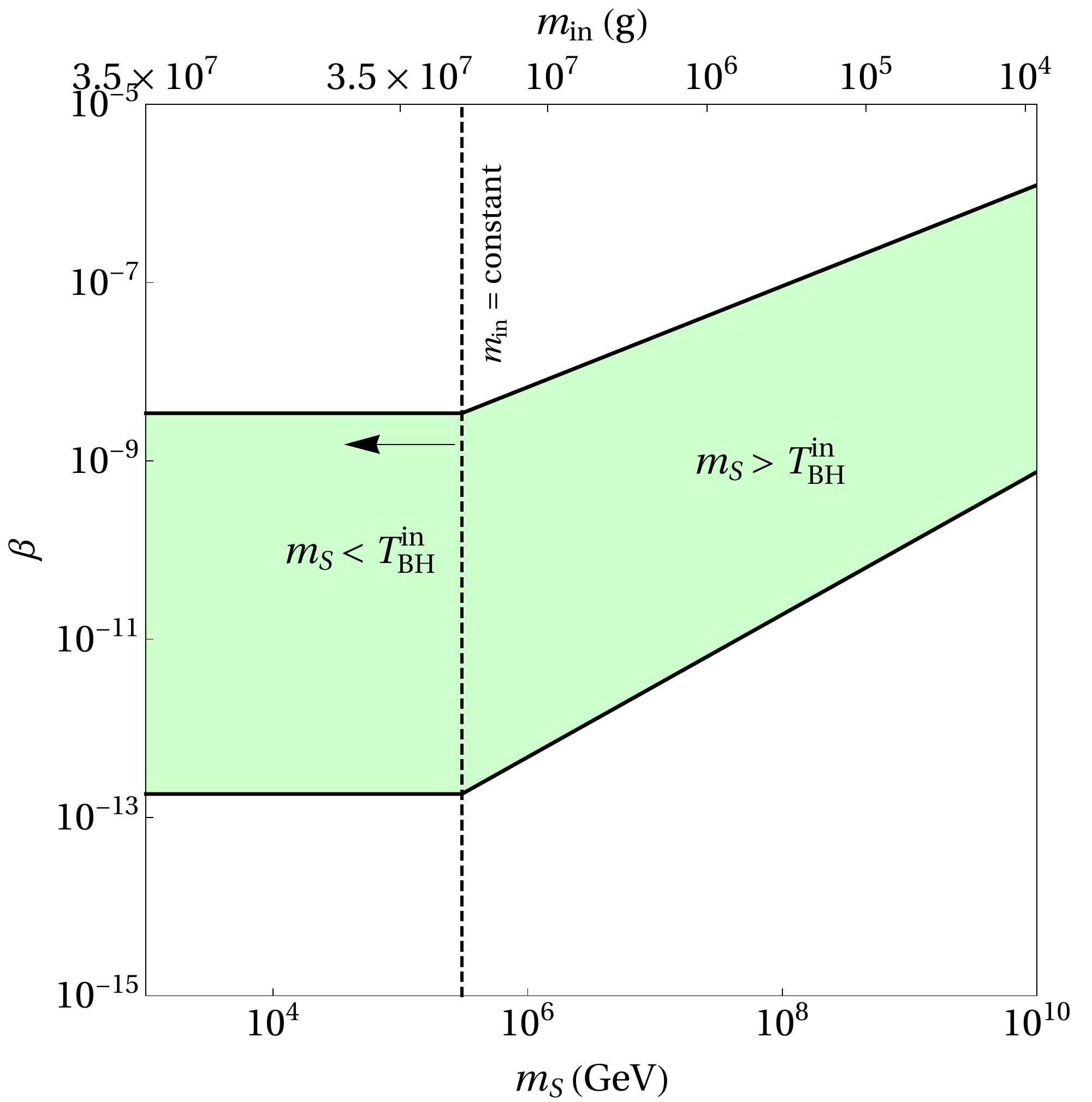}
   \caption{Bound on PBH initial fraction $\beta$ as a function of $m_S$, where $\epsilon_1+\epsilon_2 = 0.1$. For each $m_S$ we also indicate the corresponding $\Min$ in the upper horizontal axis that provides the right baryon asymmetry.}
   \label{fig:beta-yb}
\end{figure}

Once PBHs are formed, their distribution in space is random and is dictated by Poisson statistics \cite{Papanikolaou:2020qtd}. These inhomogeneities in the distribution of PBHs lead to density fluctuations which are isocurvature in nature. When PBHs dominate the energy density of the universe, the isocurvature perturbations are converted to adiabatic perturbations which at second order can induce GWs. These GWs are further enhanced due to the almost instantaneous evaporation of PBHs. The present-day amplitude of such induced GWs can be written as \cite{Domenech:2020ssp}
\begin{equation}
    \ogw(t_0,f)\simeq \ogw^{\rm peak}\left(\frac{f}{f^{\rm peak}}\right)^{11/3}\Theta
\left(f^{\rm peak}-f\right),\label{eqn:omgw}
\end{equation}
where \footnote{The total GW spectrum also has an infrared tail, which is however very much suppressed compared to the resonant contribution near the peak given by Eq.~\eqref{eqn:omgpeak} (see Ref.~\cite{Domenech:2020ssp}).} 
\begin{equation}
    \ogw^{\rm peak}\simeq 2\times 10^{-6} \left(\frac{\beta}{10^{-8}}\right)^{16/3}\left(\frac{m_{\text{in}}}{10^7 \rm g}\right)^{34/9}.\label{eqn:omgpeak}
\end{equation}
Note that the Poissonian approximation is valid only for distances larger than the mean separation between PBHs. This imposes an ultraviolet cutoff to the spectrum, with $f_{\text{peak}}$ corresponding to the comoving scale representing the mean separation of PBHs at the formation epoch.  This explains the appearance of the $\Theta$-function in Eq.~\eqref{eqn:omgw} with 
\begin{equation}
    f^{\rm peak}\simeq 1.7\times 10^3\,{\rm Hz}\,\left(\frac{m_{\text{in}}}{10^4 \rm g}\right)^{-5/6}.\label{eqn:fpk}
\end{equation}
Now, from the requirement of obtaining the observed baryon asymmetry $Y_B^\text{obs}$, one can express  $f_{\text{peak}}$ and $\ogw^{\rm peak}$ as a function of the baryogenesis scale $m_{S}$. Using Eq.~\eqref{eq:pbh-baryo} for the PBH-dominated case, we arrive at 
\begin{align}\label{eq:omgpk1}
& f^{\rm peak} \simeq 
\begin{cases}
4\times 10^{-2}\left(\epsilon_1+\epsilon_2\right)^{-5/3}~{\rm Hz} &\text{for } m_S < T_\text{BH}^\text{in}\,,\\[8pt]
2 \times 10^{-4} \left(\epsilon_1+\epsilon_2\right)^{-1/3} \left(\frac{m_S}{\rm GeV}\right)^{2/3} ~{\rm Hz} &\text{for } m_S > T_\text{BH}^\text{in}\,.
\end{cases} 
\end{align}
Similarly
\begin{align}\label{eq:fpk1}
& \ogw^{\rm peak}\simeq
\begin{cases}
7.8\times 10^{3} \left(\frac{\beta}{10^{-8}}\right)^{16/3}\left(\epsilon_1+\epsilon_2\right)^{68/9} &\text{for } m_S < T_\text{BH}^\text{in}\,,\\[8pt]
2.6 \times 10^{14} \left(\frac{\beta}{10^{-8}}\right)^{16/3} \left(\epsilon_1+\epsilon_2\right)^{68/49} \left(\frac{m_S}{\rm GeV}\right)^{-136/45}  &\text{for } m_S > T_\text{BH}^\text{in}\,.
\end{cases} 
\end{align}
From the above equations, it can be seen if the CP asymmetry parameters $\epsilon_1$, $\epsilon_2$ are fixed, the peak frequency $f_{\text{peak}}$ is determined solely by the baryogenesis scale $m_{S}$. In addition, from the requirement of obtaining the observed DM relic, $m_S$ has one-to-one correspondence with the DM mass $m_{\rm DM}$ (as seen in Fig. \ref{fig:dm}). Now, for $m_S < T_\text{BH}^\text{in}$, the frequency is degenerate for all values of $m_{S}$. On the other hand, the peak amplitude depends also on the initial PBH fraction $\beta$, which remains a free parameter in our analysis. 

In Fig.~\ref{fig:pbh-igw} , we show the GW spectra from PBH density fluctuations along with the current and future sensitivities of various GW experiments. In the left panel we fix $\epsilon_1 + \epsilon_2 = 0.1$ and as we keep on decreasing $m_{S}$, the peak frequency gets shifted to the left as expected (see Eq.~\eqref{eq:omgpk1}). For $m_S < T_\text{BH}^{\text{in}}$, the spectrum becomes independent of $m_{S}$. Similar behavior is observed in the right panel for $\epsilon_1 + \epsilon_2 = 0.01$, with an overall shift towards higher frequency. This shift  can be understood from Fig.~\ref{fig:pbh-baryo}, where the contours satisfying $Y_{B}^{\text{obs}}$ for lower CP asymmetry values requires lower values of PBH mass, and hence corresponds to a higher peak frequency (see Eq.~\eqref{eqn:fpk}). We find that for $\epsilon_1 + \epsilon_2 \lesssim 0.001$, the peak frequency becomes out of reach of any planned GW experiments shown, even with $m_S < T_\text{BH}^{\text{in}}$.  Note that in these plots we have chosen $\beta\sim \beta_{\rm max}$, where $\beta_{\rm max}$ corresponds to the upper bound given by Eq.~\eqref{eq:beta}. While this maximal value of $\beta$ is consistent with dark radiation bound at the epoch of BBN, for some choice of parameters, the peak frequencies can fall within LIGO ballpark and hence can already be disfavoured by LIGO constraints \cite{LIGOScientific:2016jlg}.

\begin{table}[htb!]
\centering
\begin{tabular}{|c | c| c| c | c| c|c|c|}
\hline
BP   & $m_S$ (GeV) & $m_{\rm DM}$ (GeV) &$\epsilon_1+\epsilon_2$ & $\Min (g)$ & $\text{log}_{10}\beta$ & $\text{log}_{10}\beta_c$ & GW experiment \\ \hline \hline
BP1 & $10^{6}$ & $3 \times 10^{12}$& 0.1 & $10^{7}$ & $-8.6$ & $-12.22$ &ET, DECIGO, BBO        \\ \hline
BP2 & $10^7$ &$2.8 \times 10^{14}$& 0.05  & $1.7 \times 10^{6}$ & $-8.2$ & $-11.45$ &CE, ET       \\ \hline
BP3 & $3\times 10^7$ &$2 \times 10^{17}$& 0.01 & $10^{5}$ & $-7.3$ & $-10.22$ &LIGO O5, CE, ET       \\ \hline
\end{tabular}
\caption{Some benchmark points (BP) of showing values of coloured scalar mass $m_S$, DM mass $m_{\rm DM}$), PBH mass fraction $\beta_{(c)}$, PBH mass $\Min$ and the CP asymmetry parameter $\epsilon_1+\epsilon_2$, along with the GW experiments that can probe the peak of the induced GW spectrum.}
\label{BP}
\end{table}

The bounds on $\beta$ given by Eq.~\eqref{eq:beta} can also be recast in terms of $m_S$, for fixed values of $\epsilon_1$, $\epsilon_2$.  This is shown in Fig. \ref{fig:beta-yb}, where we have fixed $\epsilon_1 + \epsilon_2 = 0.1$. Note that for $m_{S} <  T_\text{BH}^{\text{in}}$, $m_{\rm in}$ is fixed from $Y_{B}^{\text{obs}}$ (see Eq.~\eqref{eq:pbh-baryo} and also Fig. \ref{fig:pbh-baryo}) and hence the bounds on $\beta$ given by Eq.~\eqref{eq:beta}  become independent of $m_{S}$. The shaded region in Fig. \ref{fig:beta-yb}  represents the allowed range of $\beta$ for different values of $m_{S}$. Similar pattern is observed for different values of $\epsilon_1$, $\epsilon_2$. By keeping $\beta$ within the allowed range, we provide some benchmark values in Table \ref{BP}, along with the GW experiments which can probe the peak of the induced GW spectrum. 

\section{Conclusion}
\label{sec:concl}
We have proposed a simple way of cogenesis of baryon and dark matter from PBH evaporation which can be tested via gravitational wave induced by PBH density fluctuations. Due to the presence of new heavy particles with baryon number violation, baryon asymmetry can be generated non-thermally due to out-of-equilibrium CP violating decay of a heavy coloured scalar, predominantly produced from PBH evaporation. The allowed parameter space in terms of PBH mass consistent with the non-thermal generation of the observed baryon asymmetry also leads to the production of superheavy DM with correct relic abundance. Assuming PBH to dominate the early universe, we get one-to-one correspondence between DM mass and heavy coloured scalar mass responsible for baryogenesis. Since the induced GW spectrum also crucially depends upon PBH mass and initial fraction, we get very predictive GW spectrum at both ongoing and future experiments like LIGO, BBO, DECIGO, CE, ET etc. This is not possible in non-thermal leptogenesis from PBH scenarios where the requirement of PBH evaporation before sphaleron decoupling forces PBH mass to be much lighter leading to very high frequency induced GW out of reach from experiments. Since baryogenesis can occur at any temperature above the BBN epoch, contrary to the canonical baryogenesis via leptogenesis mechanism, one can have PBH as heavy as $\sim 10^6$ g, depending on the size of the CP-violation generated from the decay of the new coloured scalars. We find, a common parameter space, satisfying both DM abundance and baryon asymmetry can be obtained entirely from PBH  evaporation for PBHs within a mass range of $\sim 10^4-10^7$ g, where DM can be as heavy as $\sim 10^{12}$ GeV. Apart from such observable GW signatures of our cogenesis setup, the model can also have complementary signatures at collider as well as experiments searching for baryon number violation like neutron-antineutron oscillations.

\section*{Acknowledgements}
The work of RR was supported by the National Research Foundation of Korea (NRF) grant funded by the Korean government (NRF-2020R1C1C1012452). SJD would like to thank Rome Samanta for useful discussions regarding induced gravitational waves. The work of DB is supported by SERB, Government of India grant MTR/2022/000575. 

\bibliography{Bibliography, ref1, ref2}

\end{document}